\documentclass[hyper,12pt,letterpaper]{JHEP}  

\newfont{\frak}{eufm10 scaled 1200}

\newfont{\Bbb}{msbm10 scaled 1200}     
\newcommand{\mathbb}[1]{\mbox{\Bbb #1}}
\DeclareSymbolFont{AMSa}{U}{msa}{m}{n}
\DeclareSymbolFont{AMSb}{U}{msb}{m}{n}
\let\Box\relax
\DeclareMathSymbol{\Box}{\mathord}{AMSa}{"03}

\def \eqn#1#2{\begin{equation}#2\label{#1}\end{equation}} 

\def \lp{L_{p}}
\def \tr{\mbox{tr\,}}

\title{TASI Lectures on Matrix Theory}

\author{Tom Banks\\
  Department of Physics and Astronomy\\
  Rutgers University, Piscataway, NJ 08855-0849\\
E-mail: \email{banks@physics.rutgers.edu}}

\abstract{This is a summary of key issues in Matrix Theory and its
compactifications.  It is emphasized that Matrix Theory is a valid
Discrete Light Cone Quantization of M Theory with at least 6 noncompact
asymptotically flat dimensions and 16 or 32 Supersymmetry charges.
The background dependence of the quantum mechanics of M Theory, and
the necessity of working in light cone frame in asymptotically flat
spacetimes are explained in terms of the asymptotic density of states
of the theory, which follows from the Bekenstein-Hawking entropy
formula. In four noncompact dimensions one is led to expect a Hagedorn 
spectrum in light cone energy.  This suggests the possible relevance
of ``little string theories'' (LSTs) to the quantum description of four
dimensional compactifications, because one can argue that their exact
high energy spectrum has the Hagedorn form.  Some space is therefore 
devoted to a discussion of the properties of LSTs, which were first 
discovered as the proper formulation of Matrix Theory on the five torus.  
}


\received{???????? ?st, 1998}
\accepted{???????? ?th, 1998}

\preprint{\hepth{9911068}\\RUNHETC-33-99}
\begin{document}


%
\def\mth{M~Theory}
\def\matth{Matrix Theory}
\def\20n{$(2,0)_N$}
\def\two0k{$(2,0)_k$}
\section{\bf Introduction -- Limitations on Lagrangian Quantum Mechanics}

This lecture series is about Matrix Theory \cite{bfss}, 
a nonperturbative, Lagrangian formulation of 
M~Theory.  There has been a lot of confusion about this theory in the literature, 
to the extent
that it has been characterized as {\it controversial} in the popular press 
\cite{nyt}. 
Much of this confusion has been caused by misinterpretation and misunderstanding 
of what the
theory was supposed to do, and too little appreciation of how important it is to 
take the
large $N$ limit in order to calculate amplitudes of interest to a Lorentz invariant 
theory.
However, some of the difficulties of \matth\ are more important, and reflect crucial issues
about any quantum theory of gravity.  I therefore want to begin with a critical discussion
of what we can expect to be the limitations of ordinary Lagrangian quantum mechanics in the
description of {\it any} quantum theory of gravity. 

Any theory which contains general relativity (GR) must be time reparametrization invariant.
Mathematically this means that the time translation generator for an arbitrary definition of
time is a constraint, which must vanish on physical states.  Physically it means that 
any nonvanishing definition of energy must be conjugate to a physical clock variable
which measures time.   This is certain to cause problems in the quantum theory, where there must 
in general be variables which do not commute with the clock.  Thus, the very definition
of physical time translation implies some sort of semiclassical approximation in which the clock
evolves classically.  In a closed cosmology, we cannot expect such an approximation to be
valid with arbitrary precision.  However, if the universe has a boundary and is of infinite size, 
then the boundary conditions at infinity define frozen classical 
variables which can be used as clocks.
Typically, we insist that the metric at infinity approaches that of a noncompact symmetric space
(Minkowski or Anti-DeSitter (AdS)) and the natural time translation generators are chosen from
the asymptotic symmetry group of the metric.  In these lectures we will be concerned primarily
with asymptotically Minkowski spaces.

Let us first consider an ordinary Lorentz frame at spacelike infinity.  Then there is
a special Poincar\'e subgroup of the asymptotic diffeomorphisms of the
metric \cite{ashtek}, and 
up to a Lorentz transformation, a unique choice of Hamiltonian.  Quantum \mth\ (or, for those
who are still skeptical, {\it any} quantum theory of gravity) will have a Hilbert space on
which this generator acts as a Hermitian operator.  We also expect it to have a ground state 
$\vert 0\rangle$, whose
energy eigenvalue must, for consistency, be zero.  It might in fact have a discrete or
continuous ground state degeneracy, labelled by expectation values of
Poincar\'e invariant 
operators.  We will assume that, as in local field theory, there is a large class of interesting
operators (hereafter called localizable operators) 
that do not disturb the boundary conditions at infinity.  If we restrict attention
to localizable operators, the Hilbert space breaks up into superselection sectors, each of which
has a unique ground state.  Given any localizable operator $O$, we can formally define
the time dependent Heisenberg operator $O(t)$ by
\eqn{heisop}{O(t) \equiv e^{iHt} O e^{-iHt}.}
To investigate the degree of formality of this definition, we compute the two point function
\eqn{spectrep}{\langle O(t) O^{\dagger} (0) \rangle = \int_0^{\infty} dE
e^{-i
Et} \rho_O (E),}
where the spectral density is defined by
\eqn{spectdens}{\rho_O (E) \equiv \sum_n \delta (E - E_n) \left\vert
\langle 0\vert O \vert n\rangle\right\vert^2.}

The crucial question is now the convergence of this integral representation, or equivalently,
the high energy behavior of the spectral density.  In quantum field theory, the high energy 
behavior of the theory is determined by a conformally invariant fixed point.  The density
of states in volume $V$ behaves like
\eqn{confdens}{\rho \sim e^{c V (E/V)^{(d-1)\over d}},}
where $d$ is the dimension of spacetime.  Generic operators localized in the volume will
have a spectral density $\rho_O$ with the same behavior.  However, there is a special class of 
local operators of fixed dimension, which connect the vacuum only to the states in a given
irreducible representation of the conformal group.  The spectral density of these operators
grows only like a power of the energy.  Note that in either case, we can define the Green's
function by analytic continuation of an absolutely convergent integral in Euclidean time

In a quantum theory of gravity, it is extremely plausible, that for 
a theory with four or more asymptotically Minkowski dimensions the high energy density of
states is dominated by highly metastable black holes.  The existence and gross properties of these
states follow from semiclassical GR.  The density of black hole states is given by the
Bekenstein-Hawking formula
\eqn{bhdens}{\rho \sim e^{k (E/M_P)^{(d-2)\over (d-3)}},}
where $M_P$ is the $d$ dimensional Planck mass.  There are two interesting features of this
formula.  The first is its independence of the volume.  This is a consequence of the Jeans
instability.  If we try to construct an extended translation invariant state other than the vacuum
in a theory containing
gravity, we eventually get to an object whose Schwarzchild radius exceeds its physical size and the
system collapses into a black hole.  The only translation invariant states are those which are 
superpositions of a single black hole at different positions in spacetime.
Secondly, any operator whose matrix elements between the
vacuum and states of energy $E$ are not drastically cut off at energies above the Planck scale,
will not have a well defined two point function.  Thus, although the general formalism of
quantum mechanics will be valid, we should not expect a conventional Lagrangian description of
the system to be applicable.  Indeed, the Lagrangian formalism produces Green's functions
of Heisenberg operators as its fundamental output, and we are supposed to deduce the energy
spectrum and the structure of the Hilbert space from this more fundamental data (similar remarks
are of course applicable to any more abstract formalism which takes Green's functions as its
basic starting point).
In fact, implicit in the definition of the Lagrangian formalism is
an assumption that the short time behavior of the system is
approximately free.  This assumption is not even valid for a
nontrivial fixed point theory.  However, in field theory we can
restore the validity of Lagrangian methods by realizing the theory
as the limit of a cutoff system or (in many cases) by realizing
the nontrivial fixed point as an {\it infrared} limit of an
asymptotically free theory.  No such workarounds appear to
be available for the quantum theory of gravity.

Another fascinating possibility is that the space of states of a very
large black hole has a group of symmetries which partitions it into
irreducible representations in much the same way that the conformal
group partitions the states of charged black branes with AdS horizons.
Then one could construct operators which connected the vacuum only to
those states in an irrep of the group.  If the density of states
in an irrep had subexponential growth then these operators would 
have sensible correlation functions.  In the absence of such a large
black hole symmetry group, there will be no conventional Green's functions 
in the holographic dual of asymptotically flat spacetimes.

It is interesting to contrast these results with our expectations in a light cone frame.  
We will discuss light cone formalism
more extensively in the next section.  For the moment we will need only the formula for
light cone energy:
\eqn{lcen}{{\cal E} \equiv P^- = {{\bf P}^2 + M^2 \over P^+}}
where ${\bf P}$ (which we will set equal to zero) 
and $P^+$ are the transverse and longitudinal momenta respectively.  
Again assuming that the high energy density of states is dominated by black holes, we can write the
light cone density of states as
\eqn{lcdens}{\rho \sim e^{k ({\cal E} /M_P)^{(d-2)\over 2 (d-3)}}.}  
Note that we now expect a convergent two point function in light cone time for any $d\geq 5$.
Even for $d=4$, we find only a Hagedorn spectrum (rather than the more divergent form of 
the black hole spectrum in ordinary energy) and the Green's function will be defined for 
sufficiently long Euclidean time separations.  

The conclusion that I want you to draw from this is that we should only expect a conventional
Lagrangian quantum mechanics for \mth\ in light cone time, and perhaps only for $5$ or more
noncompact Minkowski dimensions.  Another point to remember is that the high energy density of
states seems to increase as the number of noncompact dimensions decreases.  This suggests that
compactified \mth\ has more fundamental degrees of freedom than uncompactified \mth\, a conclusion
which we will see is borne out in the sequel.

Some readers will be curious about the case of two or three asymptotically Minkowski dimensions,
where there are no black holes.  Here the story is quite different, at least in those situations
with enough supersymmetry (SUSY) to guarantee that there is a massless scalar field in the
supergravity (SUGRA) multiplet.  In these cases one can argue \cite{bs2d} that the system
has very few states, because would be localized excitations so distort the geometry of spacetime
that the asymptotic boundary conditions are not satisfied.  In some sense, the resulting theory
is topological.  

Another interesting example of the arguments used in this section is \mth\ in spaces with
3 or more asymptotically AdS dimensions ($AdS_2$ has the same sort of problems as two or three
dimensional Minkowski space \cite{strom}). Here the boundary of spacetime is timelike and
there is no analog of a light cone frame.  There are two natural inequivalent choices of 
Hamiltonian, corresponding to global and Poincar\'e time.  The
corresponding black objects
are AdS Schwarzchild black holes and near extremal black branes of appropriate dimension.  Both
have positive specific heat, which is to say that their density of states grows less rapidly
than an exponential.  And in precisely these cases, we expect \cite{malda} that there is
an exact description of the system in terms of a conventional quantum field theory.

In the next section, we will begin an exploration of \mth\ in the light cone frame, starting
with the simplest case of eleven noncompact directions.

\section{\bf \matth\ in Eleven Dimensions}

\subsection{Quantum Field Theory in Light Cone Frame and Discrete Light Cone Quantization}

To set the stage for our discussion of \matth\ we begin with a brief introduction to light
cone field theory \cite{ksbrodmask}.  As hinted in the previous section, the basic idea is
to choose a light cone frame pointing in a specific spatial direction.  In this basis,
the momentum takes the form $(P^-, P^+, {\bf P})$.  $P^-$ is taken to be the time translation
generator and takes the form ${\cal E} \equiv P^- = {{\bf P}^2 + M^2 \over P^+}$, where $M^2$ 
is the mass squared operator.  The subgroup of the Lorentz group which leaves the light cone
frame invariant is obviously isomorphic to the Galilean group in $d-2$ dimensions, with
$P^-$ transforming like the Galilean energy, ${\bf P}$ like the Galilean momentum, $P^+$ like
the Galilean mass, and $M^2$ like a Galilean invariant potential.  Compared to a true 
nonrelativistic field theory the new feature here is the continuous Galilean mass spectrum
and the associated
possibility of particles breaking up into others with
smaller Galilean mass.
There are also two other sets of Lorentz transformations.  
The first is the longitudinal boost generator, which rescales $P^{\pm}$ in
opposite directions.  The other is the set of null plane rotating transformations which rotate the
direction of the null vector into the transverse directions.  These are typically the most
difficult symmetries to realize in building an actual Lagrangian.

In supersymmetric theories, we must also include the supertranslations.  Since they are Lorentz
spinors, they break up into left moving and right moving pieces under longitudinal boosts.
These satisfy the following commutation relations
\eqn{susya}{[q_a , q_b ]_+ = \delta_{ab} P^+}
\eqn{susyb}{[Q_a , Q_b ]_+ = \delta_{ab} P^-}
\eqn{susyc}{[Q_a , q_b ]_+ = \gamma_{ab}^i P_i}
It will turn out in \matth\ that it is relatively easy to implement these symmetries, but that they
give strong constraints on the dynamics of the system.

For the purposes of this brief introduction to light cone 
field theory, 
we will restrict our attention to a simple scalar Lagrangian of the form
\eqn{scalag}{{\cal L} = \partial_+ \phi \partial_- \phi - ({\bf\nabla}\phi )^2 - V(\phi ).}
Standard (Dirac) quantization of this Lagrangian gives us the commutation relations
\eqn{comrel}{[\phi (z, {\bf x}, t), \partial_z \phi (z^{\prime}, {\bf x}^{\prime}, t) ] = 
{1\over 2} \delta (z - z^{\prime}) \delta ({\bf x} - {\bf x}^{\prime}).}
which are solved by
\eqn{creann}{\phi = \int_0^{\infty} {dP^+ \over P^+} [ a({\bf x}, P^+ ) e^{- i P^+ z} +
a^{\dagger} ({\bf x}, P^+ ) e^{ i P^+ z}]}
where $a$ and $a^{\dagger}$ have the commutation relations of ordinary second quantized
nonrelativistic fields.  
The $z$ independent part of $\phi$ has no canonical momentum and 
is a constraint variable.  Often one solves for it at the
classical level, but this procedure is unsatisfactory.  A better strategy (in principle) 
is to derive the light cone formalism from a covariant path integral.  Then one sees explicitly
that the zero longitudinal momentum degrees of freedom must be integrated out \cite{hellpolch} 
and that there are contributions to the effective interaction for the nonzero modes at all orders
in the loop expansion (and furthermore that
the higher order terms are larger than those obtained in
the tree approximation --- the semiclassical expansion is not applicable).

In the formalism developed so far, the zero mode problem is mixed up with an equally vexing 
problem from modes with nonzero, but arbitrarily small, longitudinal momentum.  The method
of Discrete Light Cone Quantization \cite{ksbrodmask} (DLCQ) is an attempt to repair this
difficulty by compactifying the lightlike longitudinal direction (studying the field theory
on a space where $x^-$ is identified with $x^- + R$)
, thus rendering the longitudinal
momentum discrete.  This has another property which at first sight renders DLCQ extremely
attractive.   As one can see from the expansion (\ref{creann}), the Fock
space of light cone field
theory contains only particles with positive longitudinal momentum. Operators with negative 
longitudinal momentum are annihilation operators.  If longitudinal momentum is conserved, positive
and discrete, then states with $P^+ = N/R$ can have at most $N$ particles in them.
Thus in DLCQ in the sector with $N$ units of momentum, field theory reduces to nonrelativistic
quantum mechanics with a fixed number of particles.

As there is no such thing as a free lunch, there must be a catch somewhere.  In fact, in field
theory there are two.  First of all, in order to have a hope of recapturing Lorentz invariant
physics one must take the large $N$ limit.  Field theory in a space with a periodic lightlike
direction is weird, very close to a space with periodic time, which has apparent grandfather
paradoxes\footnote{Though resolved in the way first proposed by R.A. Heinlein in \cite{zombies
bootstraps}.}.  If $N$ is large, one can hope to make wave packets which are localized along
the lightlike direction.  Furthermore, since systems with large longitudinal momenta would
be expected to Lorentz contract, their physical size in the longitudinal direction might also
be small.  The physics of such large $N$ systems could very well be oblivious to the lightlike
identification and reproduce that of the uncompactified, Lorentz invariant, system.  The use
of words like {\it might, could} and {\it hope} in the last three sentences, signals that
there is no rigorous argument guaranteeing that this is the case.

The second catch is integrating out the zero modes.  One can get some insight into how
difficult this is in field theory \cite{hellpolch} by viewing lightlike compactification
as an infinite boost limit of spacelike compactification on an infinitesimally small
circle.  If we compactify an ordinary field theory on a  circle of very small radius $R_S$ 
and concentrate
only on the lagrangian of the zero modes, then $R_S$ appears as a multiplicative factor.
In other words, the theory of the zero modes is at infinitely strong coupling.
Thus, even if the original field theory is weakly coupled, the problem of calculating
the effective Lagrangian for the nonzero modes in DLCQ appears intractable in general\footnote{
We will later encounter a field theory where DLCQ leads to a weakly
coupled system.}.

Given all of these problems, why are we interested in light cone quantization in \mth?
Apart from the general motivation given in the introduction, there are many indications
that \mth\ is much better behaved than field theory in the light cone frame.  The many successes
of light cone string theory attest to this.  In particular, in perturbative string theory in
light cone gauge, longitudinal momentum is the spatial coordinate on the string world sheet.
DLCQ is a discretization of the string world sheet.  Since the uncompactified light cone string
theory is a conformal field theory, the process of taking the large $N$ limit
is controlled by the world sheet renormalization group and the problems with zero longitudinal
momentum modes are encoded in local contact terms.  Furthermore, although the actual
computation of these counterterms to all orders in perturbation theory would be tedious, the
form of light cone string perturbation theory leads immediately to the guess that the
correct answer is given by conformal field theory on higher genus Riemann surfaces, 
a prescription which automatically fixes the contact terms by analytic continuation, at least
in many cases.

The other reason to be somewhat more hopeful about our prospects for success, is SUSY.  SUSY
nonrenormalization theorems give us some control over the possible effects of integrating out
the zero modes.  In the cases we will study, this is 
probably enough to fix the effective lagrangian
uniquely.

\subsection{The Holographic Principle and the \matth\ Lagrangian}

For a number of years, Charles Thorn \cite{thorn}
championed an approach to string theory in light cone
frame based on the notion of {\it string bits}, which were taken to be pieces of string carrying
the lowest possible value of longitudinal momentum.  Full strings, with higher values of 
longitudinal momenta were supposed to be bound states of these more fundamental constituents.
Thorn explicitly noted that in such a formalism, one of the dimensions of spacetime appeared
dynamically.  The fundamental constituents propagated on a surface of one lower dimension.

In an independent development some years later, G. 't Hooft \cite{thooft} 
proposed that the apparent
paradoxes of black hole physics in local field theory might be resolved if the fundamental
quantum theory of gravity had degrees of freedom which lived on hypersurfaces
of dimension one lower than that of the full spacetime, with a density equal to the Planck
density.  The motivation for the latter restriction was the the Bekenstein-Hawking formula
for the entropy of a black hole.
He characterized a theory of this type as {\it holographic}.  
Susskind \cite{susskind} then realized that light cone gauge string theory embodied at least
half of the holographic principle of 't Hooft, essentially because of the properties described
in the paragraph above.  The Bekenstein bound is not satisfied in perturbative string theory.
If we think of string bits as the fundamental degrees of freedom, then a string made up of $N$ bits
has a transverse extent of order $\ln N$.  On the other hand, the
Bekenstein bound would
suggest that the transverse area had to grow like $N$.  It is not terribly surprising that
this part of the holographic principle can only be realized in a nonperturbative manner in string
theory.  The Bekenstein bound is formulated in Planck units and formally goes to infinity
when the string coupling is taken to zero with the string tension fixed.  Morally speaking, it
is similar to restrictions on operators in large $N$ field theory which stem from the fact
that the traces, $\tr M^k$, of an $N\times N$ matrix are not all
independent.  It is well known
that such restrictions are nonperturbative in the $1/N$ expansion.

In nonperturbative formulations of M~Theory, such as \matth\ and the AdS/CFT correspondence,
the second half of the holographic principle is derived by explicit dynamical calculations
\cite{bfss}, \cite{sw}.  In the latter case it is rather easy to derive
and one finds that
the bound is saturated, while in \matth\ the argument is based on crude approximations and one
finds the restriction on the number of degrees of freedom only as a bound.

Let us now proceed to the construction of the \matth\ Lagrangian in eleven flat spacetime
dimensions.  The original construction of \cite{bfss} used the holographic principle as its
starting point and used the language of the Infinite Momentum Frame rather than light cone
quantization.  Susskind \cite{lendlcq} then suggested that the finite $N$ \matth\ lagrangian
was the DLCQ of \mth. Here we will follow a much cleaner argument due to Seiberg \cite{natiproof} 
(see also \cite{sen}) which begins from the idea that DLCQ of a Lorentz invariant theory
can be obtained by a boost applied to a system compactified on a spacelike
circle,  if the radius,
$R_S$, of the spacelike circle is taken to zero, with the rapidity $\omega$ of the boost scaling
like $ \ln (1/R_S )$.  If we wish to be in the sector with $N$ units of
longitudinal momentum 
in DLCQ, then we should work in the sector with $N$ units of spacelike momentum. This argument
is a derivation of Susskind's claim.

The crucial feature which distinguishes \mth\ from most field theories, is that the limiting
theory on a small spacelike circle is free.  Indeed, it is free Type IIA string theory 
\cite{witvar} (I am assuming that all the students at this school are familiar with this
paper or will shortly become so).  
The sector with $N$ units of momentum around the circle is the sector of
IIA string theory with $N$ units of D0 brane charge.  In free string theory we can
characterize this sector as containing $N$ D0 branes and the strings connecting them, as well as
any number of closed strings and D0 brane anti D0 brane pairs.  However, we are interested only
in degrees of freedom with finite light cone energy.  The light cone energy is
of order $e^{\omega}(E - N/R_S ) \sim E/R_S - N/R_S^2 $.  The lowest energy in the sector with 
$N$ D0 branes is $N/R_S$.  We are clearly interested only in states whose splitting from this
ground state is inversely proportional to the D0 brane mass.  For a single $D0$ brane,
examples of such states are the states of the D0 brane moving with (transverse from the point of
view of the 11 dimensional light cone frame) momenta fixed as $R_S \rightarrow 0$.  Note
that the eleven dimensional Planck scale remains fixed in this limit, so this is the same as 
requiring the transverse momenta to be a finite number of Planck units in the weak string
coupling limit.  For multiple D0 branes separated by distances of order the Planck scale,
we must also include degrees of freedom which create and annihilate minimal length open
strings between the branes.   The Lagrangian for this system was written down in this
context in \cite{witbound}.  It is the dimensional reduction of ten dimensional Super Yang Mills
theory on a nine torus, and, as such, 
was first written down by \cite{chetal}.   We will write it
both in string and eleven dimensional Planck units
\eqn{matlagstrng}{L = {l_S^3 \over g_S }{\rm Tr}\left({\bf
\dot{\phi}^2} +
[\phi^i, \phi^j ]^2 + i\Theta\dot\Theta
- \Theta [\phi^i, \gamma^i \Theta ] \right)}
\eqn{matlagplnk}{L = {\rm Tr}\left({{\bf \dot{X}}^2 \over R} + R{ [X^i,
X^j ]^2 \over \lp^5 } + i \theta  \dot{\theta} - R {\theta [X^i, \gamma^i
\theta] \over \lp^3}\right).}
Here $\phi^i$ and $X^i$ are nine Hermitian $N \times N$ matrices, the former
with dimensions of mass and the latter with dimensions of length.  
Similarly  $\Theta$ is a sixteen component $SO(9)$ spinor, which is
an Hermitian $N \times N$ matrix, and has dimensions of $[m]^{3/2}$,
while $\theta$ has the same transformation properties, but is dimensionless.

Witten's motivation for this Lagrangian was that it summed up the leading infrared
singularities of string perturbation theory, 
that are caused by zero energy open strings when D0 branes are separated by distances less
than the string length.  The authors of \cite{dkps} gave a careful argument {\it to all orders
in string perturbation theory} that this Lagrangian in fact captured all of the dynamics
at energy scales equal to the kinetic energy of a D0 brane with Planck momenta.
Seiberg's argument is often criticized as being ``too slick'' and the
work of
\cite{hellpolch} is cited as an example of the dangers of naively ignoring the integration
out of the zero modes in DLCQ.  In fact, the fact that \mth\ on a small circle is weakly coupled
Type IIA string theory, and the careful analysis of \cite{dkps} (which shows that the kind
of perturbative divergences of the small $R_S$ limit found by \cite{hellpolch} in field theory
are absent to all orders in perturbation theory) suggest that the latter reference is
completely irrelevant in the present context.  About the only loophole one could imagine
in the argument is the possibility that weakly coupled IIA string theory has nonperturbative
corrections to (\ref{matlagstrng}) which somehow survive the $R_S
\rightarrow 0$ limit.

Even this loophole can probably be closed by proving the following conjecture: the Lagrangian
(\ref{matlagstrng}) is the only Lagrangian {\it for this set of degrees of
freedom} consistent
with the symmetries we will list below.  The italicized phrase means in particular that
the Lagrangian may not contain time derivatives higher than the first, though it may contain
higher powers of the first time derivatives.  The conjecture has been partially proven
in \cite{chetal}.  In trying to give a more complete proof one should use certain facts
which were not employed in this reference.   In particular, the Lagrangian we have written
has a thirty two generator odd subalgebra of its symmetry algebra and has translation invariance
in the transverse directions, as well as Galilean boost invariance.  Furthermore, its gauge
group is $U(N)$ and not $SU(N)\times U(1)$, so that arbitrary separations
of the trace parts of matrices from their traceless parts are not allowed.
These facts seem to give a fairly straightforward
argument for the conjecture if one restricts attention to Lagrangians which can be written 
as a single trace.  I do not pretend to have a complete proof of this conjecture (although I
am convinced it is correct) and leave it to an enterprising student.  To my mind, the strongest 
arguments for \matth\ come from its successes in reproducing known facts and conjectures 
about \mth\ as dynamical results of a complete Lagrangian system.  There are difficult
questions about whether the large $N$ limit really reproduces the Lorentz invariant
dynamics which interests us.  But there seems to be little doubt that in a variety of backgrounds
\matth\ is a correct DLCQ of \mth.

To proceed with the exposition of the results of \matth\ we begin with a list of its symmetries:

The most important of these are SUSYs.  The full supertranslation
algebra is preserved in DLCQ.  Only the spectrum of the translation
generators is different from that expected in the uncompactified
theory.  As usual in light cone frame, spinors can be decomposed 
as right moving and left moving under the $SO(1,1)$ group of boosts
in the longitudinal direction (which is not a symmetry of DLCQ).  
Thus, there are two sets of spinor
SUSY generators, each transforming as the ${\bf 16}$ of the transverse
$SO(9)$ rotation group (which is preserved by DLCQ).  The first of
these is simply realized in terms of the $16$ canonical matrix
variables of \matth\ as
\eqn{smq}{q_a = {\sqrt{1\over R}} \tr \theta_a}
The anticommutator of these is $\delta_{ab} {N\over R}$, which 
identifies $N$ as the integer valued, positive longitudinal momentum
$P^+$ of DLCQ.
The anticommutator of the left and right moving SUSY generators is
\eqn{lrcomm}{[q_a, Q_b]_+ = {\bf \gamma P}_{ab} }
This is realized by 
\eqn{lrgq}{Q_a = \sqrt{\frac RN}\, \tr ({\bf \gamma P}_{ab}  \theta_b + i
\gamma_{ab}^{ij} [X^i, X^j ] )\theta_b }

The second term does not contribute to (\ref{lrcomm}) but is probably
required by the final relation of the supertranslation algebra
\eqn{llcom}{[ Q_a, Q_b ]_+ = \delta_{ab} P^- .}  
In fact, the latter
is realized only on $U(N)$ invariant states of the model, which
identifies $U(N)$ as a gauge group.
The word probably in the penultimate sentence reflects the
incompleteness of the uniqueness proof I referred to above.

In addition to these symmetries, the model is invariant under
$SO(9)$ rotations and transverse Galilean boosts.  The missing parts
of the eleven dimensional super-Poincar\'e group are the longitudinal
boosts and the null plane rotating parts of the spatial rotation
group.  These may be restored in the large $N$ limit.  Note that
the Galilean transformations act only on the $U(1)$ center of mass 
variables and restrict their Hamiltonian to be quadratic in canonical
momenta.

Finally, I note a discrete symmetry under $\theta \rightarrow \theta^T$,
$X^i \rightarrow - (X^i)^T$, which commutes with half of the 
supertranslations and with $P^{\pm}$.  This symmetry is instrumental
in the matrix theory description of Ho\v rava-Witten domain walls
\cite{motletal}.

\subsection{Gravitons and Their Scattering}

The classical Lagrangian of \matth\ has a moduli space consisting
of commuting matrices.  The high degree of supersymmetry of the system
guarantees that this moduli space is preserved in the quantum theory.
This means that if we integrate out all of the non moduli space
variables, then the effective Lagrangian on the moduli space has
no potential.  Furthermore, the terms quadratic in time derivatives
are not renormalized, and the terms quartic in time derivatives
appear only at one loop \cite{sethietal}.  Furthermore, for $N > 2$
there are other terms in the effective Lagrangian which receive
only a unique loop correction and thus are exactly calculable 
\cite{dinenr}.  

The justification for the description by an effective Lagrangian
is the Born-Oppenheimer approximation.  When we go off in some
moduli space direction ${\bf X_0} = \bigoplus_k {\bf r_k} I_{N_k 
\times N_k}$,  then variables which do not commute (as matrices)
with ${\bf X_0} $ have frequencies of order $\vert {\bf r_k - r_l} \vert 
$.  Thus, if these distances are large, these variables can be
safely integrated out in perturbation theory.

For the $SU(N_k)$ variables, which commute with ${\bf X_0}$ the
Born-Oppenheimer argument depends on a fundamental conjecture
about this system, due to Witten \cite{witbound}.  That is, that
the $SU(N)$ version of this supersymmetric quantum mechanics has
exactly one (up to an obvious spinor degeneracy to be discussed below)
normalizable SUSY ground state.  This conjecture lies at the heart
of the \mth\ - IIA duality, and the whole web of string dualities
would collapse if it proved false.  More impressively, the conjecture
has been more or less rigorously proven for $N=2$, and arguments
exist for higher values of $N$ \cite{sethietal}.  
Finally, arguments
can be given \cite{bfss} that the typical scale of energy of excitation
of these bound states is of order $1/N^p $ with $p < 1$.  Since we
will see that energies on the moduli space are of order $1/N$,
this justifies the use of the Born-Oppenheimer approximation for
large $N$. 

If we accept the bound state 
conjecture it follows that the large $N$ limit of the model contains
in its spectrum the Fock space of free eleven dimensional supergravitons.
In fact, the theorems we have cited show that the Lagrangian
along the moduli space direction ${\bf X_0}$ with large separations,
is
\eqn{Lag}{\sum_{k=1}^n {N_k \over 2R} \dot{\bf r_k}^2 + i \theta_k 
\dot{\theta_k} }
which is that of a collection of 
massless eleven dimensional superparticles in light cone frame.  
Each of the $\theta_k$ variables is a $16$ component $SO(9)$ spinor.
The Hamiltonian of this system is $\theta$ independent, so the
fermionic variables serve merely to parametrize the degeneracy of
particle states.  They are quantized as $16$ Clifford variables
so their representation space is $256$ dimensional.  It decomposes 
under $SO(9)$ as ${\bf 44} \oplus {\bf 84} \oplus {\bf 128}$, which are
the states of a symmetric traceless tensor, a totally antisymmetric
three tensor, and a vector spinor satisfying $\gamma^i_{ab} \psi^i_b =
0$.  This is precisely the content of the 11D SUGRA multiplet.
The required Bose or
Fermi symmetrization of multiparticle states follows from the residual
$S_n$ gauge invariance on the moduli space (commuting matrices are
diagonal matrices modulo permutations) and the fermionic nature
of the spinor coordinates.  

I want to note in particular, that SUSY was crucial to the cluster
property of these multiparticle states.  In the nonsupersymmetric
version of the matrix model, an $\vert {\bf r_k - r_l} \vert$
potential is generated on the moduli space and the whole system
collapses into a single clump.  I think that this may be one
of the most interesting results of \matth.  In perturbative string
theory, explicit SUSY breaking is usually associated with
the nonexistence of a stable, interacting vacuum state, and often
with tachyonic excitations which violate the cluster property.
\matth\ suggests even more strongly that SUSY may be crucial to the
existence of a theory of quantum gravity in which propagation
in large classical spacetimes is allowed.  In fact, it appears
that only asymptotic SUSY is strictly necessary for the cluster
property.  Indeed the cancellation of the large distance part
of the potential has to do with the SUSY degeneracy between states
at extremely high energy.  However, simple attempts to break SUSY
even softly appear to lead to disaster \cite{nss}.  

Before beginning our discussion of graviton scattering, I want to clarify what we can expect
to extract from perturbative or finite $N$ calculations in the matrix quantum mechanics.
The basic idea of the calculations that have been done is to study zero longitudinal momentum transfer
scattering by concentrating on the region of configuration space where some number of blocks are
very far away from each other.  The intra-block wave functions are taken to be the normalizable
ground states in each block.  This leaves the coefficients of the unit matrix in each block
and the off block diagonal variables.  In the indicated region of configuration space the
frequencies of the latter are very high, and one attempts to integrate them out perturbatively.
There are two rather obvious reasons why these calculations should fail to give the
answers we are interested in.  The first is that the nominal perturbation parameter for
this expansion is ${N L_P^3 \over r^3}$ where $r$ is a transverse distance between some pair
of blocks.  In order to make comparisons with SUGRA we want to take $r/L_P \gg 1$, but 
{\it independent of $N$} as $N$ tends to infinity.  Indeed the scattering amplitudes in this regime
of impact parameters should become independent of $N$ (or rather scale with very particular powers
since they refer to exactly zero longitudinal momentum transfer), as a consequence of
Lorentz invariance.   This is evidently not true for individual terms in the perturbation
expansion.  It should be emphasized that this means we are {\it not} interested in the
't Hooft limit of this theory.

In addition to this, the perturbation expansion is not even a correct asymptotic expansion
of the amplitudes in the large $r$ region.  To leading order in inverse distances, the interactions
between the high frequency off diagonal variables, and the $SU(N_i)$ variables in individual
blocks does not enter in the expressions for amplitudes.  At higher orders this is no longer
the case.  The complete calculation involves expectation values of operators in the individual
block wave functions.  The fact that the off diagonal variables have very high frequencies
allows us to make operator product expansions and limits the number of unknown expectation
values that come in at a given power of $r$.  Terms like this will give fractional
powers of the naive expansion parameter.  However, since the short time limit
of quantum mechanics is free, all the operators have integer dimensions and we are led to
expect only integer powers $L_P$ in the expansion.  The $N$ dependence of these terms is
completely unknown.  

The second of these problems is inescapable, but the first could be avoided if it were possible
to make direct comparisons between finite $N$ \matth\ and DLCQ SUGRA.  It is important to
realize that there is absolutely no reason to expect this to be so.  The intuitive reason
is that the gravitons of \matth\ are complicated bound states rather than structureless particles.
They only behave like structureless particles when their relative velocities are very small,
because then the scattering state is almost a BPS state.  In the large $N$ limit the velocities
become arbitrarily small and one can argue 
that to all orders in energies over $M_P$ they should behave like particles 
of an effective field theory.  However, for finite $N$, there is no reason for this to
be so, even at low energy and momentum transfer.  More mathematically, we can note that SUGRA
is a limit of \mth\ in which all momenta are small compared to the Planck mass.  On the other
hand, we have seen that the DLCQ system can be viewed as a system compactified on a tiny
circle, with a finite number of units of momentum.  Thus every state in the system contains
momenta large compared to the Planck scale.  There is no reason to expect the limit which
gives DLCQ to commute with the SUGRA limit.  

The reason that some amplitudes {\it are calculable} in perturbation theory is that the system
has a host of nonrenormalization theorems.  That is, the large SUSY of the \matth\ Lagrangian
so constrains certain terms in the effective Lagrangian for the relative positions
that they are given exactly by their value at some order of the loop expansion.
I will not give a description of the state of these calculations, but
refer the reader to the literature \cite{sethietal}, \cite{dinenr}.

The fact that only quantities determined by symmetries are calculable gives one pause, I must
admit (unless one is able to prove the conjecture above that the symmetries completely determine
the Lagrangian) but one must recognize that this is a rather generic state of affairs
in recent results about \mth.  However, what I consider important about \matth\ is that it
reduces all questions about \mth\ (in the backgrounds where it applies) to concrete, albeit
difficult, problems in mathematical physics.  We are no longer reduced to guesswork and speculation.
Of course, this is no better than the statement that lattice gauge theory reduces hadron
physics to a computational problem.  Obviously \matth\ will only be truly useful if one
can find analytic or numerical algorithms for efficiently extracting the S-matrix from the
Lagrangian.  On the other hand, it may be possible to attack certain conceptual problems
before a practical calculation scheme is found.  

\subsection{General Properties of the S-Matrix and the Graviton Wave 
Function}

We can easily write down an LSZ-like path integral formula for the
S-matrix of gravitons in \matth.  Simply perform the path integral
with the following boundary conditions: as $t\rightarrow - \infty$,
the matrices approach the moduli space
\eqn{lsz1}{ {\bf X} \rightarrow {\bf X_0^I} = \bigoplus_k 
{\bf r_k}(t) I_{N_k \times N_k}}
with similar formulae for the fermionic variables.  ${\bf r_k}(t)$ 
are classical solutions of the moduli space equations of motion.  They
are labelled by the transverse momenta of the incoming states, 
while the longitudinal momenta are the $N_k$.  
Similarly, for $t \rightarrow \infty$ we have
\eqn{lsz2}{ {\bf X} \rightarrow {\bf X_0^F} = U^{\dagger}\bigoplus_k 
{\bf r_k}(t) I_{N_k \times N_k} U}
where we must integrate over the $U(N)$ matrix, $U$, in order to
impose gauge invariance.  Of course, the number of 
outgoing particles, as well as their transverse and longitudinal
momenta, will in general be different from those in the initial
state \footnote{An alternate approach to the Matrix Theory S-matrix can
be found in \cite{plefka}.}.  

This formula does not quite give the S-matrix since the $SU(N_k)$ 
variables are sent to zero asymptotically by the boundary conditions.
In principle we should allow them to be free, and convolute the
path integral with the bound state wave function for each external
state.  Thus, our path integral computes the S-matrix multiplied by
a (momentum dependent) factor for each external leg equal to the bound 
state wave function at the origin.  It is likely that in the
large $N$ limit these renormalization factors will vanish, so we
would have to be careful to extract them before computing the
limiting S-matrix. 

Several of the S-matrix elements for multigraviton scattering at
zero longitudinal and small transverse momentum transfers can be
computed with the help of nonrenormalization theorems.  All of
these computations agree precisely with the formulae from 11D
SUGRA.  
As noted above, we cannot expect to make more detailed
comparisons until we understand the large $N$ limit much better
than we do at present.  

We can however try to understand what could possibly go wrong
with the limiting S-matrix.  Assuming the bound state conjecture, 
we know that the large $N$ theory has the correct relativistic
Fock space spectrum.  Furthermore, the S-matrix exists and
is unitary for every finite $N$.  This implies that individual
S-matrix elements cannot blow up in the limit.  Furthermore,
we know that some T-matrix elements are nonzero
so the S-matrix cannot approach unity (I do not have an argument
that amplitudes with nonzero longitudinal momentum transfer 
cannot all vanish in the limit).  The absence of pathological
behavior in which individual S-matrix elements oscillate
infinitely often in the large $N$ limit is more or less equivalent
to longitudinal boost invariance, which states that as the $N_k$ 
get large, S-matrix elements should only depend on their ratios.
Thus, proving the existence of generic S-matrix elements is probably 
equivalent to proving longitudinal boost invariance.  

Assuming the existence of limiting S-matrix elements, there is
another disaster that could occur in the limit.  This 
is an infrared catastrophe.  That is, the cross section for
reactions initiated by only a few particles might be dominated by
production of a number of particles scaling like a positive 
power of $N$.  Then, the S-matrix would not approach a
well defined operator in Fock space.  In 11D SUGRA the infrared
catastrophe is prevented by Lorentz invariance.  
Again we see a possible connection between the mere existence of the
S-matrix, and its Lorentz invariance. 

A possible avenue for investigating the infrared catastrophe is to exploit
the fact that the production of a large number of particles at 
fixed energy and momentum means that each of the produced particles has
smaller and smaller energy and momentum.  It is barely possible that the
nonrenormalization theorems will give us sufficient information about
scattering in this regime to put a bound on the multiparticle production amplitudes.
Personally, I believe that a demonstration of the existence and Lorentz invariance
of the limiting S-matrix must await the development of more sophisticated
tools for studying these very special large $N$ systems.

\subsection{Membranes}

One of the attractive features of \matth\ is 
the beautiful way in which  membranes are incorporated into 
its dynamics.  This connection has its origin in 
groundbreaking 
work on membrane dynamics done in the late 80s 
\cite{members}.  In that work, the \matth\ Lagrangian was
derived as a discretization of the light cone Lagrangian
for supermembranes.  The idea was to build a theory 
analogous to string theory, with membranes as the fundamental objects.
The theory appeared to fail when it was shown that the
Lagrangian had continuous spectrum \cite{dln}.  Today we
realize that this is actually a sign that the theory
exceeds its design criteria: it actually describes multibody
states of membranes and gravitons, and the continuum
states are simply the expected scattering states of
a multibody system.

The membrane/matrix connection has been described so many
times in the literature that I will only give a brief
summary of it here.  It is simplest to describe toroidal
membranes, though in principle any Riemann surface
can be treated \cite{bars}.  One of the amusing results of
this construction is that, for finite $N$, the topology 
of the membrane has no intrinsic meaning.  States describing
any higher genus surface can be found in the toroidal
construction.  It is only in the large $N$ limit that
one appears to get separate spaces of membranes with different
topology.  The question of whether topology changing
interactions (which certainly exist for finite $N$)
survive the large $N$ limit, has not been studied, but there
is no reason to presume that they do not.

The heart of the membrane construction is the famous
Von Neumann-Weyl basis for $N \times N$ matrices in
terms of unitary clock and shift operators satisfying
\eqn{clockshifta}{U^N = V^N = U^{\dagger} U = V^{\dagger}V =1.}
\eqn{clockshiftb}{UV = e^{{2\pi i \over N}} VU}
Any matrix can be expanded in a series
\eqn{expand}{A = \sum a_{mn} U^m V^n}

If, as $N \rightarrow \infty$ we restrict attention to
matrices whose coefficients $a_{mn}$ approach the
Fourier coefficents of a smooth function, $\hat{A} (p,q)$,
 on a two torus, then it is easy to show
that
\eqn{comm}{[A,B] \rightarrow {i\over N}
\{\hat{A},\hat{B} \}_{P.B.}}
and
\eqn{trace}{\tr\ A \rightarrow N \int dp dq \,\hat{A} (p,q)}
Using these equivalences, one can show that, {\it on this
subclass of large $N$ matrices}, the \matth\ Lagrangian
approaches that of the supermembrane.  

One can extend the construction to more general Riemann
surfaces (the original matrix papers cited in \cite{members}
worked on the sphere) by noting that the equations
(\ref{clockshifta}), (\ref{clockshiftb}) arise in the theory
of the lowest Landau level of electrons on a torus 
propagating in a uniform background magnetic field of 
strength proportional to $N$.  One can then study an 
analogous problem on a general Riemann surface.  Note that
since all of these Landau systems have finite dimensional
Hilbert spaces, they can be mapped into each other.  Thus,
for finite $N$, the configuration spaces of membranes
of general topology are included inside the toroidal case.

It is interesting to note that, at the level of the classical dynamics
of the matrix model, the condition (\ref{comm}) is sufficient
to guarantee that the membrane states constructed as
classical solutions of the equations of motion obeying
this restriction, will have energies of order $1/N$ and
are thus candidates for states which survive in the 
(hoped for) Lorentz invariant large $N$ limit.  This
suggests that a more general condition, {\it viz.} that
the matrices be replaced by operators whose 
commutator is trace class, may be a useful formulation of
\matth\ directly in the infinite $N$ limit.

However, it is not clear that this sort of classical 
consideration is useful when $N$ is large.  Indeed, it can
be argued that in a purely bosonic matrix model, the classical energy of
membrane states is renormalized by an amount which grows with $N$.  By
contrast, in the supersymmetric model, the infinite flat
membrane is a BPS state \cite{bss} and the energies of
large smooth membranes are all of order $1/N$ in the quantum theory. 

The direct formulation of the infinite $N$ theory in eleven
dimensions is an outstanding problem.  It is clear that it
is not simply the light cone supermembrane Lagrangian.
But perhaps supermembranes do give us a clue to the ultimate
formulation.

\subsection{Fivebranes}

We will attack the problem of finding the 5-brane of M~Theory in \matth\ by applying Seiberg's
algorithm.  In fact for longitudinal 5-branes, 
this was done by Berkooz and Douglas \cite{berkdoug} long before Seiberg's
argument was conceived of.  A longitudinal 5-brane is one which is wrapped around the
longitudinal circle.  In the IIA string language, it is an M5 brane wrapped around the small 
circle, and thus a D4 brane.  Berkooz and Douglas 
\cite{berkdoug} 
proposed that the \matth\ model for such
a 5-brane was the large N limit of the ND0-D4 system.  This is a supersymmetric quantum
mechanics obtained as the dimensional reduction of ${\cal N} =2,\ d =4$ SUSY Yang Mills, with 
an adjoint and a fundamental hypermultiplet.   For $k$ such longitudinal five branes
we simply introduce $k$ fundamental hypermultiplets.  Seiberg's argument shows that this
is the appropriate DLCQ of \mth\ with $k$ longitudinal 5-branes.
In the large $N$ limit, one can argue that a different procedure \cite{bssgrt} which dispenses
with the fundamentals, may be a sufficiently good description of the system.

We now turn to the more problematic question of fivebranes in the transverse dimensions.
After all, the longitudinal branes have infinite energy (relative to the Lorentz invariant
spectrum) in the large $N$ limit.  Again, we use Seiberg's argument and find ourselves faced
with a system containing an NS 5-brane and N D0 branes in IIA string theory with vanishing
coupling.  This is the system described by (a certain soliton sector of) 
the $k =1$ IIA {\it little string theory} \cite{dvvseiberg}.  The system
of $k$ NS 5 branes
in any string theory with vanishing coupling is a six dimensional Lorentz invariant quantum
system which ``decouples from gravity''.  That is to say, although it
contains states
with the quantum numbers of bulk gravitons (and other closed string modes) 
in a ten dimensional spacetime with a linear dilaton field \cite{maldastrom}, 
they are described holographically \cite{abkss} in terms of a quantum
theory with $5+1$ dimensional
Lorentz invariance.  It is not a quantum field theory, because it has T-duality when
compactified on circles \cite{seiberg} and because it is an interacting
theory with a
Hagedorn spectrum \cite{abks}.

The absence of a simple description of fivebranes
in the original eleven dimensional \matth\ Lagrangian is
probably the first indication of a general principle.
In quantum field theory, the fundamental degrees of
freedom are local.  When we study the theory on a compact
space we can encounter new degrees of freedom, like Wilson
lines, but they are all implicit in the local variables
which describe the dynamics in infinite flat space.
In a theory of fundamental extended objects, we may expect
that this will cease to be true.  If there are fundamental
degrees of freedom associated with objects wrapped around
nontrivial cycles of a compact manifold, and if, as may be
expected, the energies of all states associated with these
degrees of freedom scale to infinity with the volume of
the manifold, then the theory describing infinite flat space
may be missing degrees of freedom.  

Infinite fivebranes have infinite energy.  We may imagine
them to arise as limits of finite energy wrapped fivebranes
on a compact space whose volume has been taken to infinity.
Their description may involve degrees of freedom which
decouple in the infinite volume limit.  We shall see that 
this appears to be the case.  Nonetheless, one feels a
certain unease with the asymmetrical treatment of membranes
and fivebranes\footnote{Which is only partly relieved by
noting that the only true duality between the two 
\cite{oferM} {\it is} realized in the standard formulation
of \matth\ on a three torus, and that this duality {\it is} captured
by the \matth\ formulation we will present below.}.  

Perhaps there is a completely
different formulation of light cone M~Theory in which
one somehow discretizes the light cone dynamics of the
M5 brane.   Indeed, since membrane charges are certainly
incorporated in the world volume theory of the M5 brane\footnote{M5 
branes are in a sense D branes of the M2 brane
\cite{townstrom} and so carry charges which measure the
number of M2 branes ending on them.  These couple to the
two form potential on the world volume.} one might hope
to obtain a more complete formalism in this way. 

\section{\bf \mth\ on a Circle}

Let us now imagine trying to compactify one of the transverse dimensions of \mth\ on a circle of
size $ a L_P$.  We are led to study D0 branes in IIA string theory with coupling 
$g_S \sim (R_S /L_P)^{3/2} \rightarrow 0$ on a circle of radius $\sim a g_S^{1/3} l_S$.  
This situation is T dual
to $N$ D strings in Type IIB string theory with coupling $G_S \sim
g_S^{2/3}$.  The D strings 
are wound on a circle of radius $\sim g_S^{-1/3} l_S /a$.
The states of this system whose energy gap above the ground state of the D0 brane system
is of order $R_S$ are described by the $1+1$ dimensional world volume theory of nonrelativistic
D strings.  This is $1+1$ dimensional dimensional SYM theory with 16 SUSYs.  After rescaling
to light cone energy the Hamiltonian
of the system is
\eqn{matstrngham}{R \lp \int_0^{\lp^2/R_9} ds\, dt\, {\rm Tr}\left(f^2 +
({D{\bf
X} \over \lp^2})^2 + {[X^i, X^j ]^2 \over \lp^4} + \theta [\gamma D +
{\bf \Gamma X}, \theta ] \lp\right).}
In this formula, boldface characters are $SO(8)$ vectors.  
${\bf X}$ has dimensions of length.  The electric field strength $f$ has
dimensions of $[m]^2$ and $\theta $ has dimensions of mass (so that the
kinematic SUSY generator $q = {1\over\sqrt{R}} \int ds\, {\rm Tr} \theta$
has dimensions of $[m]^{1/2}$).   $R$ is the radius of the lightlike
circle.  $\gamma$ are $1+1$ dimensional Dirac matrices.  $\theta$ is a
sixteen component spinor which transforms as $(L,8_c) + (R,8_s)$ under
the Lorentz and $SO(8)$ symmetries.  The model has $(8,8)$ SUSY as a two
dimensional field theory.

It is obvious that as $R_9$ is taken to infinity, this system reduces
to the 11 dimensional matrix theory we studied in the previous section.
Indeed, the compactified theory has more degrees of freedom than the
uncompactified one.  In the Seiberg analog model, these correspond to
 strings connecting the D0 branes which wind around the torus and they
obviously become infinitely massive in the limit $R_9 \rightarrow 
\infty$.  Amusingly, in terms of the $1+1$ dimensional field theory
this decoupling is the standard one of Kaluza-Klein states when the
radius of a circle compactified field theory goes to zero.  A catchy
phrase for describing this phenomenon is that {\it in \matth\ dimensional
oxidation is T dual to dimensional reduction}.  

More interesting is the opposite limit $R_9/L_P \rightarrow 0$.  
According to the duality relation between \mth\ and IIA string theory
this limit is supposed to be the free IIA string theory.  This argument
is based on the BPS formula which shows that the IIA string tension
is the lightest scale in the theory in this limit, plus the relations
between the low energy 11D and IIA SUGRA Lagrangians.   It is important
to realize that \matth\ provides us with a true {\it derivation} of
this relation.  In a sense the relation between duality arguments and
\matth\ is similar to that between symmetry arguments based on current
algebra and the QCD Lagrangian.   

The derivation is easy.  In the indicated limit, the mass scale of
the SYM theory goes to infinity in Planck units and we should be
left with an effective conformal field theory describing any massless
degrees of freedom.  The only obvious massless degrees of freedom
are those on the moduli space, which is a $1+1$ dimensional orbifold
CFT with target space (supersymmetrized) $R^{8N} / S_N$.  This is
a classical statement, but the nonrenormalization theorems for
a field theory with sixteen SUSYs ($(8,8)$ SUSY in the language
of $1+1$ field theory) assure us that the Lagrangian on the moduli
space is not renormalized.  Indeed, we will see in a moment that 
the leading perturbation of this system consistent with the
symmetries is an irrelevant operator with dimension $(3/2, 3/2)$.

First we want to show that the spectrum of the orbifold quantum
field theory at order $P^{-} \sim 1/N$ is precisely that of the
Fock space of free Type IIA Green Schwarz string field theory
\cite{motletal}.  
To do this we note that the orbifold theory has topological
sectors not contained in the $R^{8N}$ CFT in which the diagonal
matrix fields are periodic only up to an orbifold gauge transformation
in $S_N$.  These are labelled by the conjugacy classes of the
permutation group.  A general permutation can be written as a 
product $C_{N_1} \ldots C_{N_p}$ of cyclic permutations.  Within 
each such sector, we recognize that there is a residual $Z_{N_1} \times 
\ldots Z_{N_p}$ gauge symmetry of cyclic permutation within each
block of the matrix.  

The importance of these topological sectors is that, as the $N_k 
\rightarrow\infty$, they contain states of energy $1/N_k$.  Indeed,
a diagonal matrix function on an interval of length $2\pi$,
satisfying $x_i (\sigma + 2\pi) = x_{i+1} (\sigma)
:$ mod $N_k$, is equivalent to a single function  $X_i (s)$
on an interval of length $2\pi N_k$.  The Hamiltonian for Fourier
modes of $X_i$ is $H = {1\over N_k} \sum_n n \alpha_n^i \alpha_{-n}^i$.
For a general topological sector, the Hamiltonian is a sum of $p$ such
single string Hamiltonians.  It is these {\it long strings} 
which are the strings of perturbative string theory.

There are two important constraints which follow from the remnants
of the $U(N)$ gauge symmetry of the original matrix Lagrangian.
First of all, on the subspace of states which are generated by
finite Fourier modes of the $X^i$, the $Z_{N_k}$ residual gauge 
symmetries just become translations in the variable $s$.  
In the language of string theory, these are the light cone
Virasoro constraints: $L_0 - \bar{L}_0 = 0$, on physical states.
Secondly, for configurations in which several of the long strings
are identical, there is a residual permutation gauge symmetry which
exchanges them.  This is the conventional statistics symmetry of
quantum mechanics.  It picks up the right minus signs because 
half integral spin in the model is carried by Grassmann variables.

Finally, we want to note that the single string Lagrangians
derived from this model are Type IIA Green-Schwarz superstrings. 
Indeed, the $U(N)$ SYM theory from which we began has an $SO(8)$
R symmetry group under which the left and right SUSY generators
transform as the two different spinor representations.
Thus, we may summarize the results we have derived by the statement
that {\it in the $R_{10} \rightarrow 0$, $N \rightarrow \infty$ limits,
the Hilbert space of states of the Matrix Theory Hamiltonian
with energies of order $1/N$, is precisely the Fock space of
free light cone gauge Type IIA string field theory}.
This is a derivation of the famous duality conjecture relating
M theory to Type IIA string theory.

Dijkgraaf, Verlinde and Verlinde \cite{motletal} went one step
further, and showed that the first correction to the free string
Hamiltonian for finite $R_{10}$ was an irrelevant operator which
precisely reproduced the Mandelstam \cite{mandelstam} three string
vertex.  Their argument used effective field theory: this is the
lowest dimension operator compatible with the symmetries of the
underlying SYM theory.  Thus, they were unable to compute the coefficient 
of the Mandelstam vertex.  However, because $R_{10}$ determines the
$1+1$ dimensional mass of the excitations of SYM theory which
decouple in the zero radius limit, the dimension ($(3/2 , 3/2)$) 
of the irrelevant operator determines that the string coupling
scales as $g_S \sim (R_{10} / \lp)^{3/2}$ which is the scaling
anticipated from duality considerations.  The importance of the
result is twofold.  It shows that at least to leading order in
the small radius expansion the dynamics of Matrix Theory is,
in the large $N$ limit, invariant under the ten dimensional
Super-Poincar\'e group.  And it shows that the general structure 
of string perturbation theory will follow from matrix theory.
Indeed, up to contact terms on the long string world sheet, the
Mandelstam vertex generates the Riemann surface expansion of
string perturbation theory.  As yet, no one has found an argument
that Matrix Theory provides the correct contact terms to all
orders in perturbation theory.

Let us stress the important points that we have learned in this section.  We have seen in
a very explicit way how \matth\ interpolates between the quantum mechanics of the previous section,
which describes 11D SUGRA, and free string theory.  In particular we have given a very explicit
dynamical argument for why the IIA strings are free in the zero radius limit.  In previous
discussions of the duality between IIA strings and 11D SUGRA, this was more or less a postulate,
supported only by the behavior of the low energy effective Lagrangian.  Another important lesson
was that in the limit of small radius, positions of objects on the compactified circle become
wildly fluctuating quantum variables. Indeed, these positions are
Wilson lines in the gauge theory, and we are going to the strong coupling
limit.
 There is no longer any sensible geometrical meaning to 
the small circle, but the theory itself is perfectly smooth in the limit.

\section{\bf \mth\ on a Two Torus}

To study the theory on $T^2$ we apply the same set of arguments.  D0 branes in weakly coupled
Type IIA on an 11D Planck scale torus are treated by double T duality and related to D2 branes 
in a dual (but still weakly coupled) Type IIA theory.  The states of finite light cone energy
are described by maximally symmetric $2+1$ dimension SYM theory, with finite coupling,  
compactified on a torus which is dual to the \mth\ torus.  (In fact, here is an exercise: go through
Seiberg's argument for a general torus and show that the states of finite light cone energy
are those whose energy scale is given by the coupling of the SYM theory obtained by doing T duality
on all the radii.  For $T^4$ and up this SYM theory is nonrenormalizable and we will see the 
implications of this in the next section.)  The two Wilson lines of the SYM theory represent
the coordinates of particles on the \mth\ torus.   

Aspinwall and Schwarz \cite{aspschw} argued using string duality that Type IIB theory in ten
dimensional space was obtained as the zero area limit of \mth\ on a two torus.  The seeming
contradiction that $11 - 2 = 10$ is resolved by noting that in the zero area limit a continuum
of light wrapped M2 brane states appear, and play the role of momentum in a new tenth dimension.
One of the puzzles of this approach is why the theory should be symmetric under rotations which
rotate this new dimension into the other $9$.  Fundamental (F) and D strings are identified as
M2 branes wrapped around the short resp. long cycle of the torus.  This identifies the type
IIB string coupling as the ratio of the short and long cycles (more generally, the imaginary part
of the complex structure), explains the $SL(2,Z)$ duality
of the theory and explains why there is a T duality between weakly coupled IIA and IIB theories.

If we try to take the zero area limit of the \mth\ torus we are led to study the SYM theory
on a torus of infinite area.  Again, because the SYM coupling is relevant, this is equivalent
to an infinite coupling limit in which all but conformal degrees of freedom decouple.
Again, the existence of a moduli space ensures us that there is some sort of conformal
limit rather than a completely trivial topological theory.  Now however, there is a difference.
In $2+1$ dimensions there is a finite superconformal algebra which has 16 ordinary 
supercharges.  It has an $SO(8)$ R subalgebra under which the supercharges
transform as 
(8,2) (with the 2 standing for their transformation under the $2+1$ dimensional Lorentz group).
Furthermore, since the theory is conformal, it depends only on the complex structure of the
limiting $T^2$ and there is an obvious $SL(2,Z)$ invariance which acts on
the complex structure
\cite{bs}.

The finiteness of the superconformal algebra allows for the possibility of an interacting
theory, and indeed the same interacting superconformal theory was postulated to describe
the interactions of $N$ M2 branes at separations much smaller than the 11D Planck scale.
Here we want the theory to be interacting because on a two torus with finite complex structure 
we are trying to describe interacting Type IIB string theory.
We can understand the weak coupling limit, and simultaneously get a better understanding
of the $SO(8)$ symmetry, by going to large complex structure.  In the
limit where the $a$
cycle of the \mth\ torus is much smaller than the $b$ cycle, the
corresponding SYM torus has
cycles with the opposite ratio.  Thus we can do a Kaluza Klein reduction on the cycle of the
dual torus corresponding to the $b$ cycle and obtain a $1+1$ dimension
field theory.  

This is best done by going to the moduli space, which is a $U(1)^N$ SYM theory before the Aspinwall
Schwarz limit.  To take the limit we do an electromagnetic duality transformation, replacing
the Abelian gauge fields by compact scalars ($F_{\mu\nu} \propto
\epsilon_{\mu\nu\lambda}\partial^{\lambda} X^8$), 
which decompactify in the limit.  It is then obvious that
there is an $SO(8)$ symmetry which rotates this boson into the seven
original scalars.  
Indeed the Lagrangian on the moduli space is 
\eqn{2dmodlag}{{\cal L} = (\partial_{\mu} X^i )^2 +
\bar{\theta_a}\Gamma^{\mu} \partial_{\mu} \theta_a }
Here $i$ and $a$ each run from $1$ to $8$, and the $\theta$'s are
two component, $2+1$ dimensional spinors.  Of course, $2+1$ Lorentz
invariance is broken by the compactification on a torus, but under the 
$SL(2,Z)$ which transforms the radii of the torus, the spinors transform
as a doublet.  There is an obvious $SO(8)$ symmetry of this Lagrangian
under which the $X^i$ and $\theta_a$ could each transform in any of the
eight dimensional representations.  Note that both components of
$\theta_a$ must transform the same way.  Superconformal invariance in
the Aspinwall Schwarz limit, assures us that this $SO(8)$ group survives
in the interacting theory.  Furthermore it tells us that the scalars
must transform in the vector representation of $SO(8)$ and the fermions
in one of the two chiral spinors (which one is a matter of pure
convention).
                                                              
When we further make the Kaluza-Klein reduction corresponding to large complex structure
the moduli space theory is identical to the light cone gauge IIB Green-Schwarz string.
Actually it is $N$ copies of this theory, related by a residual $S_N$ gauge symmetry 
(as in the previous section).  We can rerun the analysis there and show that the theory
is the Fock space of free strings
in this limit and that the first correction to the limit is the correct Mandelstam
interaction for IIB strings, with the right scaling of the couplings.  The
$SO(8)$ symmetry
is seen to be the spacetime rotation symmetry of IIB string theory, and superconformal
invariance in $2+1$ dimensions has given us an understanding of the reason for emergence of this
symmetry (without recourse to a weak coupling expansion) and of the chirality of the resulting
spacetime physics.  An alternative derivation of the $SO(8)$ symmetry
using the compactification
of the theory on the three torus can be found in \cite{ss}.   

One thing which does {\it not} work is the continuous 
perturbative shift symmetry of the theory under
translations of the Ramond-Ramond scalar.  This can be attributed to the virtual presence in
the theory of various longitudinally wrapped branes which are sensitive to the value of
$\theta$.  Presumably this will all go away in the large $N$ limit when the energies of
these states go off to infinity (relative to the Lorentz invariant states).  
This example show us that, at least in perturbation theory there can be many versions of
DLCQ.  A perturbative DLCQ of Type IIB string theory would have preserved the continuous 
symmetry.   

\section{\bf Three and Four Tori}

The theory on the three torus is somewhat less interesting.  One again obtains a maximally
SUSY YM theory, which is scale invariant and has an $SL(2,Z)$ Olive
Montonen duality symmetry.
This combines with the geometrical $SL(3,Z)$ to give the proper U duality 
of \mth. There are now no new limits.  Olive Montonen duality is identified with the M duality
\cite{oferM} which identifies M2 branes with M5 branes wrapped on a three torus, giving a new
version of \mth.  
The four torus is much more interesting.  Going through the Seiberg scaling one obtains
D0 branes in weakly coupled IIA theory on a Planck scale four torus.  Performing four T dualities
to get to a a large manifold we come back to IIA theory, but this time with a large coupling
$G_S$ (because the coupling rescales by four powers of $g_S^{-1/3}$.  The zero branes have of 
course become D4 branes.  Large coupling means that we are going back to \mth\, a new dimension is
opening up, and the D4 branes become M5 branes.  We are thus led to the theory of $N$ M5 branes
at distances well below the Planck length in a new copy of \mth.  
This is a $5+1$ dimensional superconformally invariant field theory
$(2,0)_N$ \cite{seibwtiseib16}
whose moduli space is $N$ self dual tensor multiplets with Wilson surfaces lying in the
self dual $U(N)$ weight lattice.  This proposal was first made by
\cite{rozali}.
In terms of the original
IIA theory we can understand the extra momentum quantum number in the field theory
as arising from D4 branes of the (pre T duality) IIA theory wrapped around the Planck scale
torus.  

The $(2,0)_N$ theory is superconformal and lives on a five torus.  The smallest radius 
of this torus defines the Planck length of \mth\ and the remaining four torus is the
dual torus to the one \mth\ theory lives on.   The $SL(5,Z)$ U duality of \mth\ compactified
on $T^4$ is manifest in this presentation of the theory.

Govindarajan and Berkooz and Rozali \cite{berkroz} 
have suggested that a very similar construction can be made
for \mth\ compactified on K3 manifolds.  One first uses the Seiberg
scaling to relate the problem to D0 branes on a K3 of scale $g_s^{1/3} l_S$ in the zero coupling
limit of IIA string theory.  Then one uses the K3 T duality 
to relate this to D4 branes wrapped on a dual K3 in a strongly coupled IIA string theory
and thus to the \20n theory compactified on $S^1 \times \hat{K3}$.  They show that many properties
of the theory, including all of the expected dualities and the F theory \cite{vafaf} limit
can be qualitatively understood in this formulation.

\section{\bf Five and Six (Where We Run Out of Tricks)}

We now come to the five torus, where things really start to get interesting.
The standard limiting procedure leads us to D0 branes in weakly coupled IIA string
theory on a Planck scale 5 torus, which is T dual to D5 branes wrapped on a 5 torus
in strongly coupled IIB string theory, which in turn is S dual to NS 5 branes wrapped
on a five torus in weakly coupled IIB string theory.  If one goes through the dualities
carefully, then one sees that the scale that is being held fixed in the latter theory,
is the IIB string scale.  We can see this as follows.  As usual, in the original picture,
we are taking the DKPS limit and the scale which is held fixed is the kinetic energy
of a single D0 brane with Planck scale momenta.  After T duality this always leads at
low enough energy to a SYM theory in which the SYM coupling is held fixed.  In 
Type IIB theory, with string coupling $G_S$, the SYM coupling is given by
\eqn{weakIIB}{g_{SYM}^2 = G_S L_S^2}
Let us rewrite this in terms of the parameters of the S dual IIB theory: 
\eqn{sduala}{\tilde{G_S} = 1/G_S,}
\eqn{sdualb}{\tilde{L_S}^2  = G_S L_S^2. }
 We learn that a collection of $N$ coincident NS 5 branes in weakly
coupled IIB string theory, has, on its collective world volume, a SYM theory whose
coupling depends only on the string tension and does not go to zero with the string coupling.
As a consequence we learn (believing always in the consistency of string theory) that the
$G_S = 0$ limit of a collection of $N$ coincident NS 5 branes in the zero coupling limit
of Type IIB string theory is a consistent interacting quantum theory with manifest $5+1$
dimensional Lorentz invariance.  We call this the $U(N)_B$ little string theory.
The Matrix Theory for \mth\ on a five torus with 
Planck scale radii and $N$ units of longitudinal momentum is the $U(N)_B$ little string theory
compactified on a dual five torus.  The parameter 
 $\tilde{L_S}$ and radii $\Sigma^A$ of the little string theory are related to those of \mth\ by 
\eqn{lstma}{1 = {R^2 L_1 L_2 L_3 L_4 L_5 \over L_S^2 \lp^9}}
\eqn{lstmb}{\Sigma^A = {\lp^3 \over R L_A}.}
Here $R$ is the lightlike compactification radius.

The little string theory retains the manifest $O(5,5)$ T duality of compactified IIB string
theory with zero coupling.   This is now interpreted as the duality group of \mth\ and is
in fact the correct U duality group of \mth\ on a five torus.  This symmetry has several
interesting consequences.    First of all, the little string theory cannot be a quantum
field theory.  In quantum field theory, the variation of correlation functions with respect
to the metric is given uniquely in terms of insertions of the stress tensor.  But T-duality
transformations change the metric of the torus without changing the theory.  If we make a small variation
of a radius of the torus around some T-self dual
point, then we find (assuming that we are dealing
with a quantum field theory) that every matrix
element of the operator $\int \delta g^{\mu\nu} 
\theta_{\mu\nu}$ vanishes.   It is easy to argue
that this is incompatible with the properties
of field theory.    We will find more evidence
below that little string theories are not field
theories.

Another consequence of T duality is the existence
of another type of little string theory, called
$U(N)_A$.  Indeed if we do a T duality 
transformation on a single radius of the five torus,
we get $N$ NS five branes in IIA theory.  If we now
take the infinite torus limit, we obtain a distinct
theory.   This could have been obtained directly
by considering the zero coupling limit of NS five
branes in the IIA theory in infinite ten dimensional
spacetime.   Indeed, one can construct similar
little string theories from the zero coupling limit
of NS five branes in the heterotic theories.
The low energy limit of the $U(N)_A$ little string theory is not a SYM
theory but rather the \20n superconformal field 
theory.  

An interesting way of understanding the fact that
interactions of NS five branes survive the limit
of zero string coupling is to write down the
low energy SUGRA solution corresponding to a
collection of $N$ NS fivebranes.  In the string
conformal frame this has the form
\eqn{ns5grav}{g_{ij} = \delta_{ij} e^{2\phi}}
\eqn{ns5dil}{e^{2\phi} = e^{2\phi_0} + {N\over r^2}}
\eqn{ns5tor}{H_{ijk} = N v_{ijk}}
Here the indices span the four dimensional space
transverse to the five brane and $r$ is the Euclidean
distance from the five brane in this space.  
$v_{ijk}$ is the volume form of the unit three
sphere in this space.  $e^{2\phi_0} = 
{\tilde{G_S}}^2$, the
square of the string coupling.  The metric components
along the fivebrane are Minkowskian.

Near $r=0$ the transverse space has the form of a flat infinite
one dimensional space times a three sphere of fixed radius.  The
dilaton varies linearly in this flat space.
This limiting background is in fact an exact solution
of the classical string equations of motion to
all orders in $\alpha^{\prime}$ \cite{CHS}.  
Indeed, the 
background $H$ field converts the three sphere
$\sigma$ model into the level $N$ $SU(2)$ Wess-Zumino-Witten (WZW)
conformal field theory.  The linear dilaton is
such that the value of the super central charge
is $\hat{c} = 10$.  This is called the linear
dilaton background.

For finite $\tilde{G_S}$ this infinite space is cut off on
one end and merges smoothly into an asymptotically
flat space with finite string coupling.  However,
no matter how small the asymptotic value of the 
string coupling, the region near the five brane
is strongly coupled.  The effect of taking 
$\tilde{G_S}$ 
to zero is to make the linear dilaton background
valid everywhere.  As a consequence of the fact
that the coupling goes to infinity at $r=0$,
the higher order terms in the formal genus
expansion are infinite and the perturbation 
expansion is not useful for correlation functions
which probe the $r=0$ region.  We will see later
that for large $N$, duality allows us to study
this region in terms of a different low energy
SUGRA expansion.

In addition to the failure of quantum field theory
to capture the dynamics of DLCQ \mth\ on $T^5$,
this geometry presents us with another new 
phenomenon.  In all previous cases, Seiberg's
argument allowed us to relate DLCQ \mth\ to
string theory, but we then found that most of
the string states decoupled in the DLCQ limit.
Here we have found that the scale which is kept
finite is the string scale and we are left with
a little string theory.  Another name we might
have chosen for this is a ``Kondo string
theory''.  The famous Kondo model in
condensed matter physics is a free $1+1$ dimensional
field theory interacting with a localized
defect with a finite number of degrees of 
freedom.  The full system is a rather nontrivial
interacting quantum problem, and the 
degrees of freedom of the field theory cannot be 
thrown away even though they are free everywhere
except at the position of the defect.

Maldacena and Strominger \cite{maldastrom} have argued that
the same is true for the little string theory.
If we analyze the scattering problem of string
modes off a five brane in the zero coupling limit
it is easy to convince oneself that every 
asymptotic state of the string theory maps into
an asymptotic state of the linear dilaton background.
Maldacena and Strominger argue that at large $N$
 any state of
the fivebrane with energy 
density\footnote{``All the experts'' agree that the same 
conclusions are valid for localized states of
finite energy on the fivebrane, although no 
calculations have been done for this case.  The
case of finite energy density is directly relevant
to the toroidally compactified little string theory
which is our essential concern in Matrix Theory.} 
above a certain cutoff can be described in a sufficiently
good approximation as a $1+1$ dimensional black
hole in the low energy effective field theory.
The point is that for large enough $N$ and energy density, the 
black hole horizon is in the region where the
coupling is still weak.  Thus, the classic
Hawking analysis of black hole radiation is valid
and indicates that such fivebrane states decay
into asymptotic string states.
A standard calculation shows that the Hawking
temperature is of order 
$1 / \sqrt{N\tilde{L_S}^2}$.
This has two interesting consequences.  First
it shows that in the large $N$ limit, 
there {\it is} a true decoupling of these string
states.  Second, since the temperature is independent
of the energy, it implies a Hagedorn spectrum
of black hole states, which must be interpreted
as states of the little string theory.  
This is a second indication that little string 
theories are not field theories.  We will find
independent confirmation of this spectrum by
a very different method below.

The large $N$ decoupling of the string states
is extremely important, because it is easy to
see that in the \matth\ context these states
do not have Lorentz invariant dispersion relations.
First of all, since their D0 brane charge vanishes,
they have no longitudinal momentum.  Secondly,
in the DLCQ limit they actually have vanishing
transverse momentum as well.  Indeed, the typical
states to which perturbative string theory applies 
have
transverse momenta of order the string scale as
the string coupling goes to zero.
The weakly coupled string theory which we use
in the description of \mth\ on $T^5$ is an
S-dual Type IIB theory, whose string length
in terms of the original Type IIB theory
is given by equation (\ref{sdualb}).  In turn
we have, in terms of the original IIA string coupling
$G_S = o(g_S^{- 5/3})$.  Thus, a momentum of
order $\tilde{L}_S^{-1}$ is of order 
$g_S^{5/6} L_S^{-1}$.  By contrast, the D0 branes
have momenta of order the eleven dimensional Planck
scale, which scales like $g_S^{-1/3} L_S^{-1}$.
Thus, in the limit, string states of the little
string theory carry zero Planck units of \mth\ 
transverse momentum. 

This remark explains the otherwise paradoxical
fact that the transverse momentum of string states
in the NS 5 brane background is not conserved
(note that the $O(4)$ {\it angular momentum} is
conserved).
From the point of view of the flat space string 
theory we began from, the string modes are 
interacting with states which carry infinitely
more transverse momentum than they do, and therefore
they can gain or lose arbitrary amounts of
(string scale) transverse momentum.

From the \mth\ point of view then, the string states
of the little string theory are troublesome.
They carry finite light cone energy but exactly
zero transverse and longitudinal momentum.
They are not consistent with \mth\ Lorentz invariance.
Fortunately, they seem to decouple in the large $N$
limit.  The Maldacena Strominger 
calculation seems to indicate that excitations
on the five branes do not excite such states 
(even if it were energetically possible) in the
large $N$ limit \footnote{O.Aharony has argued to me that since the
spectrum of stringy states appears to begin only at energies of orde
$1/\sqrt{N}$ times the string scale, there is an energetic argument for
decoupling, independent of the Maldacena-Strominger calculation.}.

On the six torus, things get even more out of hand.
After performing the Seiberg limit and using T 
duality we obtain the theory of D6 branes in a
strongly coupled Type IIA string theory.  We
are instructed to keep the SYM coupling on the
D6 brane world volume finite.  It is well known,
that in this limit, D6 branes can be viewed as
KK monopoles of 11D SUGRA compactified on a 
very large circle.  Be very careful to note
that this is {\it not} the 11D SUGRA we are
trying to model.  In fact, the gravitons of this
11D SUGRA arise from the IIA DLCQ point of view,
as D6 branes wrapped on the \mth\ $6$ torus.
The SYM coupling on the KK
monopole world volume is just the Planck scale
of this (new) 11D SUGRA.

This is a new wrinkle.  Previously the theories
which described DLCQ \mth\ did not contain gravity.
This was an advance because the conceptual problems
of quantizing gravity seemed to be avoided.
This is no longer the case on $T^6$.  The only
saving grace here is that one can again argue that
these fake gravitons had better decouple in the
large $N$ limit.  Indeed, the reader may verify
that, just like the string states of the little
string theories, they carry vanishing longitudinal
and transverse momenta from the \mth\ point of
view.  This means they had better decouple.
A hand waving argument that they do in fact decouple
is the following.

KK monopoles are manifolds which are circle bundles
over the space transverse to a six brane.  The radius
of the circle is fixed at infinity (though we must
take the limit in which this asymptotic radius
is itself infinite) and goes to zero near the
six brane.   For a monopole of charge $N$ the rate
at which the circle shrinks to zero as the radius
is varied, is multiplied by $N$.
 Thus gravitons with nonzero momentum\footnote{It is
easy to see that gravitons with zero momentum
 decouple in the limit that the SUGRA circle goes to 
infinity.} around the
circle will be repelled from the KK monopoles, and
the repulsion will set in at a larger distance
for large $N$.  From the point of view of \mth\
we want to study the scattering of $N$ KK monopoles
(wrapped on the dual 11D SUGRA $T^6$) at transverse
separations much smaller than the dual Planck scale
(although we want to keep energies which are of
order the dual Planck scale).  
It seems plausible that these scattering processes
will not involve graviton emission in the large
$N$ limit.  Obviously, we could do with a stronger
argument.

The example of $T^6$ kills once and for all the idea
that the finite $N$ DLCQ should reduce to finite $N$
DLCQ SUGRA in the limit of low energy and large
transverse separations.  It is clear that at finite
$N$, DLCQ \mth\ contains states of arbitrarily low
light cone energy (wrapped D6 branes in the original
description -- gravitons in the T dual description)
which are simply not there in DLCQ SUGRA.

One might have thought that the simple scaling
arguments above go through for any compactification
on a six manifold.  However, Seiberg's argument
implicitly contains assumptions about the moduli
space of string theory compactified on manifolds
smaller than the string scale --- assumptions which
are valid only if there is enough SUSY to provide
nonrenormalization theorems for the space and the
metric on it.  The argument indeed goes through
for $K3 \times T^2$ but the authors of \cite{kls}
have pointed out that things are quite different
for a general Calabi-Yau threefold, where there
are only eight supercharges.  Indeed, it is
well known \cite{wittencandetal} that the K\"ahler
moduli space of string theory on 
CY 3-folds is corrected when 
the sizes of cycles reach the string scale.
The exact form of the K\"ahler moduli space
and the metric on it can be read off from
the complex structure moduli space of the
mirror manifold \cite{agmetal}.  The authors
of \cite{kls} suggest that the point in
moduli space corresponding to a ``Planck
scale Calabi Yau'' is a mirror CY whose
complex structure is very close to the conifold 
point.   This conjecture is based on the
notion that mirror symmetry is obtained \cite{pmsyz}
by writing the CY as a $T^3$ fibration
and doing T duality on the three torus.  It is not
precisely clear what this means since the manifold 
has no Killing vectors with which to perform an
honest T duality transformation.  Nonetheless,
the idea that mirror symmetry would map 
a very small (real) 6 fold into a 6 fold with
a shrinking three cycle sounds plausible.

If this suggestion is correct, then we know at least
that the effective theory for the DLCQ 
will not contain gravity.  Indeed, it is known
since the seminal work of Strominger \cite{andycon}
that the effective theory of the new massless states
coming from wrapping Type IIB three branes on
the shrinking 3 cycle, is a $3+1$ dimensional
gauge field theory with a massless hypermultiplet.
The authors of \cite{kls} suggest that the whole
Matrix Theory on a CY threefold may be some sort
of $3+1$ dimensional field theory with four
supercharges.  This is an interesting idea, but
not much follow up work has been done on it.  In
my opinion it is a direction which may lead to
some interesting progress.

We have seen that Matrix Theory becomes more and
more complicated as we compactify more and more
dimensions.   This is quite interesting, since
it is not the way field theory behaves.  When we
compactify a field theory we generally lose degrees
of freedom rather than gain them\footnote{To make 
this statement more precise, count the number of
degrees of freedom below a certain energy, and ask
how this number changes as we shrink the size of
the compactification manifold.}.
This is not
completely true.  In gauge field theory 
compactification adds Wilson lines, and in gravity,
it adds the moduli of the compactification manifold.
However this addition is far outweighed by the loss
of modes with nontrivial variation on 
small manifolds.  Perhaps more importantly, these modes
did exist as gauge degrees of freedom on the noncompact manifold,
but with gauge functions which cannot live on the
compactified space.

Some of the extra degrees of freedom we have discovered in
\matth\ are artifacts
of DLCQ.  In the low energy SYM approximation,
the momentum modes of the field theory represent
(from the original \mth\ point of view) branes wrapped
around both transverse and longitudinal cycles.
These states have energy of order one when
 $N\rightarrow\infty$ and should decouple from the
hypothetical Lorentz invariant limiting theory.

Examples where this can be worked out rather explicitly are
the weak coupling limits of Type II \cite{motletal} and Heterotic
\cite{banksmotletal} strings, as derived from various
$1+1$ and $2+1$ SYM theories.  There it is seen that only
certain quasi-topological modes of the SYM theory, which vary at
a rate $1/N$ along the SYM torus (and manage to be periodic
by wandering a distance of order $N$ in the space of matrices),
survive the large $N$ limit.  In my opinion, the key question
in the dynamics of \matth\ is to find a way to isolate and 
describe the spectrum of order $1/N$ with an effective
Lagrangian, away from the weak coupling limit.  

\subsection{The Seven Torus and Beyond}

From the point of view described in the introduction, the problems we have
encountered as we increased the number of compactified dimensions
beyond four are connected to the density of states of the
theory at large energies.  The little string theory has, as
we shall see below, a Hagedorn spectrum.  This is the essential
feature that prevents it from being a quantum field theory.
The DLCQ of \mth\ on a six torus does not decouple from gravity.
As a consequence, its light cone density of states grows faster than
an exponential, because its high energy 
light cone spectrum is identical to
that of SUGRA in an ordinary reference frame.

As we have emphasized, these problems should go away in the large
$N$ limit.  The Lorentz invariant spectral density
 of the models grows more slowly than an exponential.  Indeed, for
both the five and six tori we have suggested that the offending
states decouple in the large $N$ limit.   

On the seven torus we face a problem of a somewhat different nature.
It has long been known that massive excitations of a Lorentz invariant
vacuum in $2+1$ dimensional gravity do not preserve globally
asymptotically flat boundary conditions.  Worse, in theories with
massless scalar fields in $2+1$ dimensions (which includes SUGRA
with all but the minimal SUSY) excitations tend to have 
logarithmically growing scalar Coulomb fields and infinite energy.
This has been argued to imply \cite{bs2d} that the Hilbert space
of Lorentz invariant, 
asymptotically flat $1+1$ or $2+1$ dimensional string theory
is topological in nature and contains no local propagating
excitations.

In DLCQ we compactify one more dimension than necessary to
describe the Lorentz invariant system we are trying to model.
Thus, the paucity of states with asymptotically flat boundary conditions 
should become a problem in compactifications to four spacetime
dimensions.  Indeed, following Seiberg's argument for \matth\ on
the seven torus we are led to a theory of seven branes in Type IIB
string theory.  The BPS formula tells us that these have logarithmically
divergent tension.  Thus, there is no sensible DLCQ of \mth\ with
$1+1$, $2+1$, or $3+1$  dimensional Lorentz invariant asymptotics.
Note that we do expect a noncompact formulation of light cone \mth\
with $3+1$ dimensional asymptotics to exist (it should have a Hagedorn
spectrum, like little string theory), but it cannot be found
as the large $N$ limit of DLCQ.

\section{\bf DLCQ and Holography 
of \two0k Theories and Little String Theories}

In our discussion of compactification of \matth\
we encountered two new types of Lorentz invariant
quantum theories which seemed to be decoupled from
gravity in the sense that they could be formulated
on fixed spacetime manifolds.  This is certainly
true for the \two0k theories and their less 
supersymmetric cousins, which are ordinary
quantum field theories in six dimensions.
It is likely to be true for the little string 
theories as well.  

In this section we will introduce two complementary
methods for studying these theories.  At the moment,
both methods make sense only in flat six dimensional
Minkowski spacetime.  Even toroidal compactification
results in new singularities which are 
not well understood.  Although we could treat the
field theories as limits of the little string 
theories we will instead find it useful to introduce
both methods of computation in the simpler
context of field theory.  We begin with DLCQ.

\subsection{DLCQ of \two0k Theories}

We have remarked above that DLCQ is not a terribly
useful tool for ordinary field theory because the
theory compactified on a small spatial circle is
usually strongly coupled and intractable.  This
is not the case for the \two0k theories.  Indeed,
dimensional reduction on a small circle leads
us to an infrared free $4+1$ dimensional SYM theory.

One simple argument for this comes from the 
derivation of the \two0k theory as the effective
theory of $k$ coincident M5 branes.  If we compactify
on a small spatial circle along the brane
then we are studying $k$ coincident D4 branes
in weakly coupled Type IIA theory.  Things become
even simpler if we ask what in the SYM theory
corresponds to momentum around the small circle.
The only obvious conserved quantum number is
the instanton number (remember that instantons
are particles in $4+1$ dimensions).  That this
is indeed the right identification follows from
the BPS formula for the instanton mass 
$M_I = 8\pi^2 /g_{SYM}^2$.  Remembering the 
identification of the coupling in terms of the radius
of the fifth dimension, we see that this
is just the formula for the mass of a KK mode,
$M_{KK} = 2\pi /R_5$.  

Since the SYM coupling is small when the radius is
small, and the $4+1$ dimensional SYM theory is
infrared free, a semiclassical analysis of the
dynamics of the instantons is valid.
Thus, in the sector with longitudinal momentum $N$,
DLCQ of the \two0k theory would seem to reduce 
to quantum mechanics on the moduli space of $N$
instantons in $U(k)$ gauge theory.  The fact that
this moduli space and the quantum mechanics on it
are calculable from classical considerations follows
from the high degree of SUSY of the problem\footnote{All 
of these arguments come from the
papers \cite{abkss} and \cite{abs}, while the 
regularization below was invented in the second
of these two papers.}.

Well, almost.  The fly in the ointment is that this
moduli space is singular.  Fortunately, there is
an elegant and unique regularization of the moduli
space of instantons in four Euclidean dimensions,
which appears to make the system completely finite
and sensible.   The \two0k theory has 16 ordinary 
SUSYs.  In light cone frame we expect only half of
them to be realized linearly so we expect to find
a quantum mechanics with 8 SUSYs.  The target space
of the quantum mechanics must therefore be a 
hyperk\"ahler quotient.   There is a famous 
construction (called the ADHM construction) \cite{adhm} of instanton moduli
space as a singular hyperk\"ahler quotient.
It is the solution space of
the algebraic equations
\eqn{adhma}{ [X,X^{\dagger}] - [Y,Y^{\dagger}] + q_iq_i^{\dagger} - (p^i)^{\dagger} p^i =0}
and
\eqn{adhmb}{[X,Y] = q_i p^i}
modded out by a $U(N)$ gauge symmetry which acts
on  $X$ and $Y$ as adjoints and the $k$ $q_i$ and $k$ $p^i$ as fundamentals
and antifundamentals respectively.
The products of fundamentals and adjoints appearing in these equations
are tensor products of $U(N)$ representations and are to be
interpreted as matrices in the adjoint representation.

  These equations also define
the Higgs branch of the moduli space of ${\cal N} 
= 2,\ d=4$ $U(N)$ SYM theory with 
$k$ fundamental hypermultiplets.  The latter
interpretation also introduces the natural
regularization of the space, for we can add
a Fayet Iliopoulos term by modifying the first ADHM equation to read
\eqn{adhmfi}{ [X,X^{\dagger}] - [Y,Y^{\dagger}] + q_iq_i^{\dagger} - (p^i)^{\dagger} p^i = \zeta I_N,}
where $I_N$ is the $N\times N$ unit matrix, and $\zeta $ is a real number.
Note that this is a regularization of the moduli
space but not of the Yang Mills equations as local
differential equations.  Instead it corresponds to
solving the Yang Mills equations on a certain
noncommutative geometry \cite{ns}.

An important facet of this DLCQ of the \two0k theories is
the fact that when they are KK reduced on a circle, the 
low energy effective theory is five dimensional SYM theory,
which is infrared free.  Thus, the difficulties encountered
in \cite{hellpolch} should be absent and the semiclassical
identification of the system as quantum mechanics on the ADHM
moduli space is valid.  SUSY nonrenormalization theorems
guarantee that the metric on this space is unique, and
the regularization of the singularities by the FI term is
the unique way to deform the instanton moduli space into a 
smooth hyperk\"ahler manifold.  
The key to finding the spectrum of chiral primary operators 
in the \two0k theory from DLCQ is the following observation of
\cite{abs}.  These authors observe that the DLCQ procedure 
preserves a subgroup of the superconformal group of the full theory.
They identify these generators as  explicit operators in the
quantum mechanics on instanton moduli space.  In particular, they 
show that
a vertex operator is primary, only if it is concentrated
on the singular submanifold of zero scale size instantons.
Chiral primary operators can then be identified in terms
of the cohomology with compact support of the Fayet-Iliopoulos
regulated instanton moduli space, which has been investigated
in the mathematical literature.  We will not explore
the details of these calculations.  Suffice it to say that
they find the correct spectrum of chiral primary operators.

The way we know this is that the spectrum calculated from DLCQ
coincides with that implied by the AdS/CFT correspondence.  This is
not just a matter of agreement between two unrelated conjectures
(which in itself would be impressive).  Rather, the basis for the
AdS/CFT identification of primary operators comes from a low energy
analysis of the interaction of 11D SUGRA with fivebranes.  There
must be one primary  for each SUGRA field which is in a short
multiplet of $AdS_7 \times S^4$ SUSY.   As usual with short
multiplets, the number and properties of these multiplets
are independent of parameters, and can be calculated in the low energy
approximation.

\subsection{DLCQ of the Little String Theories}

The papers \cite{abkss} and \cite{withiggs} 
described the DLCQ of the $U(k)_A$ little
string theories and \cite{sethgan} performed
the same task for the $U(k)_B$ theories.  We will
restrict attention to the $U(k)_A$ case.

One way to understand the derivation for the Type
A theory is to consider the DLCQ of Type IIA
string theory as derived in the Matrix String 
picture and add fivebranes wrapped around the
longitudinal direction.  The result is the
$1+1$ dimensional field theoretical generalization
of the model of Berkooz and Douglas \cite{berkdoug}
for longitudinal 5 branes in Matrix-Theory.
One obtains a $1+1$ dimensional field theory with
$(2,2)$ SUSY.  It is a $U(N)$
gauge theory with one adjoint
and (in the sector with $k$ fivebranes) $k$ 
fundamentals.   As in \cite{motletal}
one takes the weak string coupling limit by
descending to the moduli space.   Now however we
want to be on the Higgs branch of the moduli
space (the Higgs and Coulomb branches obviously
decouple from each other in the limit) and we
obtain a $\sigma $ model with target space the
ADHM moduli space.  
Of course this moduli space is singular, but
we can regularize it by adding FI terms.  Thus,
the DLCQ${}_N$ of $U(k)_A$ LST is a sigma model on the moduli
space of $N$ $U(k)$ instantons on $R^4$.  This moduli space
can be regularized by the addition of FI terms to the ADHM
equations, so that the DLCQ theory is realized as a limit of
a well defined, conformally invariant $(4,4)$ supersymmetric
sigma model.  Note that, in contrast to the regularized 
quantum mechanics,
the sigma model retains its conformal invariance after regularization.
However, the conformal generators of the sigma model are {\it not}
symmetries of the spacetime \mth.  From the \mth\ point of view, the
spatial momentum on the sigma model world sheet is a quantum number
that counts longitudinally wrapped branes, and should decouple
in the limit of large $N$.

Before regularization the ADHM moduli space
is locally flat.  Thus, the central charge $c$ of the SCFT
is just $c= 6Nk$.  Because the variation of the FI term is
a marginal perturbation of the sigma model, this remains the value
of $c$ in the regularized model.  We can immediately turn this
into a computation of the high energy density of states in 
the DLCQ model.  The entropy is given by
\eqn{sdlcq}{S(P^-) \rightarrow \sqrt{2 c P^- } =\sqrt{ 6k} E l_S}
In the last equality we have used the relation between light
cone and ordinary energy for vanishing transverse momentum.
This is the Hagedorn spectrum that we advertised for the little 
string theories.  In the DLCQ approach, it arises, as in perturbative
string theory, because the light cone energy is identified with
the ordinary energy of a $1+1$ dimensional CFT.

The only problem with this derivation is that it applies to the
asymptotic density of states of the DLCQ theory, which are not
actually states with $P^{-}$ of order $1/N$.  It has been argued by
Ofer Aharony \cite{oferpriv} that the sigma model contains such
states as strings which wander through of order $N$ instantons before
closing (note that the instanton moduli space has an $S_N$ orbifold
symmetry).  As in our treatment of matrix string theory, or the
Maldacena-Susskind description of fat black holes \cite{maldasuss},
these configurations should have an entropy of the Hagedorn form
even at energies much lower than those at which (\ref{sdlcq}) is
naively valid in the CFT.  We will see below that a completely different
argument produces the same formula for the entropy.
 
The sigma model on regularized instanton moduli space is a fascinating
CFT, which also arises in the study of D1-D5 black holes.  Its properties
have recently been studied in \cite{swdab}.

\subsection{Holography}

The AdS/CFT correspondence will be covered by
other lecturers at this school.  Suffice it
to say that for the \20n theory\footnote{In the previous section, in
order to conform to the literature on the subject
we used $k$ to signify the number of five branes, while $N$ was reserved
for the longitudinal momentum.  Here we will revert to the standard use
of $N$ in the AdS/CFT correspondence.  It denotes the number of fivebranes.} 
it provides
the leading term in a large $N$ expansion of
the correlation functions of the chiral primary
operators.  The calculation is performed by solving
the classical equations of 11D SUGRA in the
presence of certain perturbations of an $AdS_7 
\times S^4$ spacetime, with $N$ units of fivebrane
flux on the $S^4$.  In order to calculate
corrections to higher order in $1/N$ one needs
the higher terms in the derivative expansion
of the effective action.  We have only a limited
amount of information about such terms.
Even in leading order, few calculations have been
done for this case.

Just like the DLCQ solution of these theories,
the AdS/CFT calculations only work in 
uncompactified spacetime.  The SUGRA background
corresponding to toroidally compactified \20n
theories is singular and the derivative expansion
does not seem sensible even for large $N$.  
In this case we can get an inkling 
of the reason for the greater
degree of singularity of the compactified case.
We are used to the fact that in \mth, singularities
correspond to light degrees of freedom which have
not been included in the effective Lagrangian.
In the toroidally compactified \20n theory it is
obvious that there are such degrees of freedom.
The zero modes along the moduli space, which are
frozen expectation values in the infinite volume
theory are here zero frequency quantum variables.
Indeed it is precisely the scattering matrix of
these variables which one would hope to compute
in \matth.  It seems unlikely that this dynamics
will be easily captured by reliable calculations
in SUGRA.   

We now turn to the holographic description of
little string theories.  It was suggested in
\cite{abks} that these are simply the exact
description of string theory in the linear dilaton
backgrounds.  More precisely, these are Type II
string theories in the following background.
\eqn{metric} {ds^2 = -dt^2 + d{\bf x}^2 + d\phi^2  + a N l_s^2 
d\Omega_3^2 }
\eqn{ast}{H = \omega_{N \Omega_3}}
\eqn{dil}{g_S^2 = e^{- \phi/\sqrt{N} l_S}}
Here ${\bf x}$ are coordinates on the worldvolume of $N$ coincident
fivebranes, $\Omega_3$ are coordinates on a three sphere transverse
to the fivebranes, and $\omega_{\Omega_3}$ is its volume form.

There are some interesting differences
in the way holography works in this context, as
compared to the AdS/CFT correspondence.  Most of
them stem from the fact that the asymptotic geometry
of these backgrounds is Minkowski space with an
exponentially vanishing string coupling.  Therefore,
even for finite $N$, there is an infinite region
of spacetime in which the description of the system
in terms of freely propagating particles becomes
exact.  The linear dilaton systems have an S-matrix.
By contrast, even though the geometry of AdS space has
infinite volume, the boundary conditions which define 
a Cauchy problem in this space are reminiscent of
those for a system in a box.   

In the AdS/CFT correspondence we do not expect
to see any sort of large spacetime unless $N$
is large, but even for $N=1$ (note that unlike
other systems of this type, the $N=1$ 
little string theory does not appear to be a 
trivial gaussian system)
 or $2$ the little
string theories should have asymptotic 
multiparticle states propagating in the weak
coupling region.  

Another dramatic difference between the two
types of theories is that the little string theory
has a Hagedorn spectrum and is not a quantum
field theory.  Thus, in many ways, the 
little string theory is much closer to string
theory in Minkowski space than the AdS systems.

We have seen the Hagedorn spectrum in the DLCQ calculation
above.  In the holographic description one calculates
the asymptotic density of states by using the Bekenstein-Hawking
formula for black hole entropy.  This is justified in the
linear dilaton background because the mass of the black hole
is inversely proportional to the string coupling at the horizon.
The world outside a large mass black hole is completely
contained in the weak coupling regime.  

It is well known \cite{cghs} that the Hawking temperature of 
linear dilaton black holes is independent of their mass.  This
is equivalent to the statement that the entropy is linear in the
energy {\it i.e.} we have a Hagedorn spectrum\footnote{It should
be noted in passing that this simple calculation shows that all 
extant nonperturbative formulations of the $c=1$ string theory
are wrong.  The entropy in all such calculations is that of
a $1+1$ dimensional field theory rather than the much more 
degenerate Hagedorn spectrum.  The $c=1$ model was solved by
trying to resum a divergent perturbation expansion.  Clearly some
nonperturbative states (Liouville ``D branes''??)
have been missed in this resummation.}.  The Hagedorn temperature
can be computed in terms of the cofficient which governs the rate
of increase of the dilaton.  We again find that  $S = \sqrt{6N} E l_S$.

The Hagedorn spectrum actually solves a potential paradox in the 
claim of \cite{abks}.  These authors argue that the S-matrix of
string theory in the linear dilaton background can be interpreted
as the correlation functions of observables in the LST.  The $p$ particle
S-matrix elements are of course symmetric under interchange of
arguments (the $S_p$ group of statistics).  They are also
$5+1$ Lorentz invariant.
If these are to
be interpreted as correlation functions in a quantum theory, their
$S_p$ symmetry implies that they must be Fourier transforms of
time ordered products of Heisenberg operators.

Lorentz invariance would then imply that LST was a local field
theory, because time ordered products are only Lorentz invariant
if the operators commute at spacelike separations.  
However, the Hagedorn spectrum prevents us from performing
the Fourier transform and this conclusion cannot be reached\footnote{
Some readers may be confused by our apparent denial of the 
possibility of having a Hagedorn spectrum in local field theory.
What, they will ask, about the Hagedorn spectrum of large $N$ QCD?
In fact there is no contradiction.  Two point functions of operators
in large N QCD are in fact controlled by the asymptotically free
fixed point at short distances.  However, the crossover scale,
above which free behavior sets in, depends on the operator.  At
infinite $N$ there are some operators which never get to the
crossover point, because it scales with a positive power of $N$.
This phenomenon of operators which have rapidly vanishing matrix
elements between the vacuum and most of the high energy states, 
appears to be connected to the
fact that infinite $N$ QCD is a free theory, with an infinite
number of conservation laws.  I do not expect such behavior in
a finite system with interaction.}.

Indeed, by calculating two point functions of operators by analogy
with AdS/CFT (solving linearized wave equations in the linear
dilaton background), Peet and Polchinski \cite{pp} showed explicitly
that they were not Fourier transformable.  Their behavior is
exponential, with the Hagedorn temperature controlling the rate
of growth of the exponent.

Peet and Polchinski's calculation is easy to summarize:
The scalar wave equation in the linear dilaton background is
\eqn{lindilscal}{[-\partial_{\phi}^2 + {(2l+3)(2l+1)\over 4} + k^2
\alpha^{\prime} N] e^{3\phi/2} \psi = 0.}

For large $k$ and $\phi$ its solutions have the form
\eqn{lindilsoln}{ \psi \sim e^{{(\alpha^{\prime} N k^2)}^{1/2}}\phi}
In the holographic interpretation the S-matrix computed from these
wave functions, which has the same behavior in momentum space as the
wave functions themselves, is supposed to be the two point function
of some operator in the LST.  Thus, it the two point function is not
Fourier transformable, and grows in the way we would expect from 
the Hagedorn spectrum. 

To conclude this brief summary of our knowledge of little 
string theories,
I want to discuss the question of what the scale of nonlocality is
in these theories.  What we know so far suggests two rather different
answers.  Seiberg's original argument about T-duality suggests
string nonlocality on a scale $l_S$.  On the other hand, the length
scale defined by the Hagedorn temperature is of order $l_S \sqrt{N}$
which is much longer.  Note however that the argument for the latter
scale is based on high energy asymptotics.  Thus, although the
Hagedorn temperature is low for large $N$, it might be that the
exponential behavior of the density of states does not set in
until energies of order $l_S^{-1}$.  The Hagedorn temperature
controls the rate of growth of the asymptotic density of states,
but does not tell us anything about the finite scale at which
the asymptotic behavior begins to dominate.  Minwalla and Seiberg
have done a calculation which suggests that in fact the Hagedorn
behavior does not set in until scales far above the Hagedorn
temperature.  They argued that if, in the $LST_A$
 theory, one takes the limit $l_S \rightarrow
0$ with $l_S \sqrt{N}$ fixed, then the SUGRA approximation to
scattering amplitudes with energies of order this fixed scale
becomes exact.  The point is that in $LST_A$, the strong coupling
behavior of the theory is described at low energy by 11D SUGRA.  
Minwalla and Seiberg show that in the large N limit described
above, there is a SUGRA description of the scattering amplitude
which is valid for arbitrary values of the dilaton.  Thus,
the full amplitude is calculable by solving partial differential
equations.  The resulting equation is complicated, but Minwalla
and Seiberg obtained a qualitative understanding of its behavior
and were able to solve it approximately in various regimes.

They calculated the amplitude for a single massless
string to scatter off the NS 5 brane in this limit, and found 
a Fourier transformable answer.  This suggests that for large $N$,
the density of states in the vicinity of the Hagedorn energy scale, 
increases more slowly than the Hagedorn formula\footnote{This is not 
a definitive argument against a Hagedorn spectrum because the matrix 
elements of operators between the vacuum and high energy spectrum
might fall sufficiently rapidly to give a Fourier transformable two
point function.  It is however suggestive that the limiting
Minwalla-Seiberg calculation shows a different behavior than that found
by Peet and Polchinski.}.
It is tempting to suggest that the Hagedorn behavior of the
spectrum sets in only above the string scale, which is the scale
of nonlocality indicated by T duality.  Indeed, in the spacetime
picture of this system, the high energy CGHS black hole spectrum
can only be computed reliably for energies above the string scale.
If this conjecture is correct, there is a puzzle about
the nature of the large $N$ limiting theory defined by Minwalla and
Seiberg.  Naive
application of the logic we applied to the full LST would suggest
that it is a quantum field theory, since its correlation functions
have spacetime Fourier transforms, which can then be 
interpreted as Lorentz invariant time ordered products.  
But large $N$ limits are tricky, and I expect that if the 
Minwalla-Seiberg limit of all the correlation functions of $LST_A$
exists, it does not define a quantum field theory.

Finally, I want to discuss an issue raised by the analysis of
Minwalla and Seiberg, which is not particularly related to the bulk
of the material in these lectures.  There is some confusion in the
literature, and in discussions I have participated in,
about whether the AdS/CFT correspondence (and in particular the
fact that the theory is formulated in terms of a Hermitian
Hamiltonian in a well defined Hilbert space) says something
definitive about the issue of unitarity in Hawking radiation.
I would claim that it does not, because the AdS theory does
not have an S-matrix\footnote{Attempts to extract the flat
space S-matrix from AdS/CFT, \cite{polchsuss}, have not progressed
to the point where one can decide if the S-matrix is unitary.}.  

The little string theories do have an S-matrix and one can begin
to address the question.  In particular, Minwalla and Seiberg
find a nonzero absorption cross section for the black fivebrane.
This could be taken as a signal of lack of unitarity.  Like
many extremal black holes, the extremal fivebrane metric has
an analytic completion with multiple asymptotic regions.  One
could try to interpret the absorption cross section as matter
being scattered into another asymptotic region, violating unitarity
in any given region.  

I would like to present a more conservative interpretation of 
the absorption cross section: the fivebrane absorbs only because
it is infinite.  There is indeed another asymptotic region,
but this is the region along the infinite brane.  This is most clearly
seen in the IIB case, where the low energy theory on the brane
is infrared free $5+1$ dimensional SYM theory.  A particle
coming in from infinity in $\phi$, can excite gluons on the brane 
which propagate off to asymptotic infinity in $5$ dimensions without
any propagation back off to infinite values of $\phi$.  
In the IIA theory analyzed by Minwalla and Seiberg \cite{shiraz}, 
the low energy
theory is conformal and has no conventional S-matrix.  However
it should still be true that localized disturbances in CFT
eventually dissipate out to infinity\footnote{I thank A.Zamolodchikov
for a discussion of this point.}, so similar physics is to be
expected.  

The key test of this interpretation is to see what happens when the 
theory is compactified.  Indeed, the origin of the absorption
cross section is a logarithm in momentum space which comes from the
continuum of low energy modes.  Since these are described by field
theory, we understand what happens to them upon compactification.
The modes on the brane are discrete.  There will be no absorption
at generic energies.  All that will happen is the excitation of
resonant modes of the brane, which will eventually decay back to the 
vacuum.  The only continuum in the system is that describing
modes propagating far from the brane, in the weak coupling
region.  

On the gravitational side of the holographic correspondence,
compactification causes a singularity to appear on the horizon
(the compactification circles have zero radius there).
One cannot conclude anything from this about unitarity, but
it certainly does not contradict the conclusion based on the
nongravitational side of the duality.  Thus, there is no evidence
against unitarity of the LST S-matrix.

Little string theories are a fascinating area for future work.  They
are our only example of Lorentz invariant quantum theories which are
neither quantum field theories
nor theories of gravity (in $5+1$ dimensions).  Conventional
Lagrangian techniques are applicable only in the light cone frame.
It would be of the utmost interest to find an alternative, manifestly
Lorentz invariant, framework for formulating and solving these
theories.  

\section{\bf Conclusions}

Matrix theory is a nonperturbative DLCQ formulation of \mth\ in 
backgrounds with six or more asymptotically flat directions.
It provides proofs of a large number of duality conjectures,
and has led to a new class of Lorentz invariant, gravity free
theories.  It demonstrates the existence of a new class of
large $N$ limits of ordinary gauge field theories, in which
one concentrates on states with energies of order $1/N$.
There is a lot of evidence that the theory becomes simpler in the
large $N$ limit, in the sense that many of the finite $N$ degrees
of freedom decouple.  A Lorentz invariant formulation
awaits the development of techniques to study these new kinds
of large $N$ limit.

In the meantime, we can try to use \matth\ to study a variety of
issues in gravitational physics which do not require us to
compactify to low dimensions.  A beginning of the study of black
holes in \matth\ may be found in \cite{matbh}.

There are a number of important general lessons about \mth\ that may
be learned from \matth.  Among these are

\begin{enumerate}
\item The statistical gauge symmetry of identical particles arises
as a subgroup of a much larger, continuous, gauge symmetry.

\item The cluster property, and the existence of spacetime itself
seems to be closely intertwined with supersymmetric cancellations.

\item The number of degrees of freedom of the theory increases
as we compactify.  This is quite odd from the point of view of
quantum field theory.  

\item Short distance divergences in the effective gravitational
theory turn out to be infrared divergences caused by the 
neglect of degrees of freedom which become light when particles
are brought together.  These correspond to light branes stretched
between the particles, and again are very different from the
kinds of degrees of freedom encountered in field theory.

\item As in any generally covariant theory we expect a conventional
Hamiltonian description only when space is asymptotically flat or
AdS.  In the asymptotically flat case we have argued that conventional
Hamiltonian quantum mechanics will only be applicable in the light
cone frame and only when there are five or more noncompact
dimensions.  The phenomenologically relevant case of four dimensions
has a Hagedorn spectrum in light cone energy and may be describable
by some kind of little string theory.

\end{enumerate}

The outstanding problem in \matth\ is to find a way to isolate the
dynamics of the states with DLCQ energy $1/N$ and to write a Lagrangian
(for $d_{noncompact} \geq 5$) for the infinite $N$ system.  For the
phenomenologically relevant case of $d=4$ one must obtain a sensible
substitute for Lagrangian methods for systems with a Hagedorn spectrum.
Another unsolved problem is the formulation of DLCQ \mth\ on Calabi-Yau
threefolds. Beyond this, \matth\ cannot go, for light cone methods do
not appear to be useful for cosmology or for studying the problem of
SUSY breaking (where the typical ground state of the system may not have
null Killing vectors).   

\acknowledgments

I am grateful to the organizers, J.Harvey, S.Kachru, E.Silverstein, 
and especially K.T.Ma\-han\-tap\-pa for inviting me to this stimulating
school. This work was supported in part by the DOE under grant
number DE-FG02-96ER40559.


\end{document}